\documentclass[%
 reprint,
superscriptaddress,
 amsmath,amssymb, aps,
]{revtex4-2}

\usepackage{graphicx}
\usepackage{dcolumn}
\usepackage{bm}
\usepackage{float}
\usepackage{xcolor}
\usepackage{subfig}
\usepackage[hidelinks]{hyperref}
\usepackage{chemformula}
\usepackage{makecell}
\usepackage[justification=justified,singlelinecheck=false]{caption} 

\newcommand{\subfigimg}[3][,]{%
  \setbox1=\hbox{\includegraphics[#1]{#3}}
  \leavevmode\rlap{\usebox1}
  \rlap{\hspace*{10pt}\raisebox{\dimexpr\ht1-1\baselineskip}{#2}}
  \phantom{\usebox1}
}
\begin{document}

\preprint{APS/123-QED}

\title{Effects of Strong Capacitive Coupling Between Meta-Atoms in rf SQUID Metamaterials}

\author{Jingnan Cai}
\email{To whom all correspondence should be addressed. E-mail: jcai@umd.edu}
\affiliation{Quantum Materials Center, Physics Department, University of Maryland, College Park, MD 20742-4111 USA}

\author{Robin Cantor}
\affiliation{STAR Cryoelectronics, 25-A Bisbee Court, Santa Fe, NM 87508-1338 USA}

\author{Johanne Hizanidis}
\affiliation{Institute of Electronic Structure and Laser, Foundation for Research and Technology-Hellas \& Department of Physics, University of Crete, 70013 Herakleio, Greece}

\author{Nikos Lazarides}
\affiliation{Academic Support Department, Abu Dhabi Polytechnic, P.O. Box 111499, Abu Dhabi, United Arab Emirates}

\author{Steven M. Anlage}
\affiliation{Quantum Materials Center, Physics Department, University of Maryland, College Park, MD 20742-4111 USA}



\date{\today}

\begin{abstract}

We consider, for the first time, the effects of strong capacitive and inductive coupling between radio frequency Superconducting Quantum Interference Devices (rf SQUIDs) in an overlapping metamaterial geometry when driven by rf flux at and near their self-resonant frequencies. The equations of motion for the gauge-invariant phases on the Josephson junctions in each SQUID are set up and solved. \textcolor{black}{Our model accounts for} the high-frequency displacement currents through capacitive overlap between the wiring of SQUID loops. We begin by modeling two overlapping SQUIDs and studying the response in both the linear and nonlinear high-frequency driving limits.  By exploring a sequence of more and more complicated arrays, the formalism is eventually extended to the $N\times N \times 2$ overlapping metamaterial array, where we develop an understanding of the many ($8N^2-8N+3$) resulting resonant modes in terms of three classes of resonances. The capacitive coupling gives rise to qualitatively new self-resonant responses of rf SQUID metamaterials, and is demonstrated through analytical theory, numerical modeling, and experiment in the 10-30 GHz range on capacitively and inductively coupled rf SQUID metamaterials.
\end{abstract}

\maketitle


\section{\label{sec:intro}Introduction}

 A Radio Frequency (rf) Superconducting Quantum Interference Device (SQUID) is simply a superconducting 
loop interrupted by a single Josephson junction (JJ). This device was originally introduced to measure small dc magnetic fields and to operate as a magnetic-flux-to-frequency transducer \cite{Silver67,Hansma73,RifkinrfS76,Muck01}. Later, it was recognized that rf SQUIDs, \textcolor{black}{due to their low loss, self resonance, and strong interaction with electromagnetic fields, can be used as meta-atoms in} a metamaterial, both in the quantum \cite{Du06,Rak08}, and classical limits \cite{Laz07,Mam10}. The rf SQUID effectively acts as a macroscopic-quantum split-ring resonator.  \textcolor{black}{The rf SQUID metamaterials were initially realized by covering a plane with rf SQUIDs to act as a nonlinear and highly tunable metasurface} \cite{Butz13,Jung13,Trep13}.  

The properties of rf SQUID metamaterials can be tuned by means of dc magnetic flux $\Phi_\text{dc}$, rf magnetic flux $\Phi_\text{rf}$, temperature, and the structure of the metamaterial, which dictates the interactions between the meta-atoms. The closed superconducting loop creates the condition for magnetic flux quantization, while the single Josephson junction in the loop introduces the Josephson inductance $L_\text{JJ}=\frac{\Phi_0}{2\pi I_c(T) \cos \delta}$ \cite{Anlage11,Jung14,Laz18}. Here $\Phi_0=h/2e$ is the magnetic flux quantum, with $h$ Planck's constant and $e$ the electronic charge, $I_c$ is the critical current of the junction, and $\delta$ is the gauge-invariant phase difference across the junction. The self resonant frequency of the rf SQUID is dictated by the junction capacitance and shunt capacitance (which are in parallel and have a total capacitance $C$) along with the geometrical inductance of the loop $L$, and the Josephson inductance $L_\text{JJ}$.  A dc magnetic flux applied to the SQUID loop creates screening currents that adjust the gauge-invariant phase $\delta$ in the JJ, and thus tune $L_\text{JJ}$ through a wide range of positive and negative values \cite{Chris71,Rifkin76}.  The result is a highly tunable self resonant frequency that can span the mm-wave, microwave, and RF frequency ranges \cite{Butz13,Jung13,Trep13}.  In non-hysteretic SQUIDs ($\beta_\text{rf} \equiv 2\pi L I_c/\Phi_0 < 1$) the flux tuning is periodic with period $\Phi_0$.  The application of rf flux to the SQUID modifies the impedance of the JJ through the Josephson effect nonlinearity \cite{Aur73}, $I=I_c(T)\sin \delta$, \textcolor{black}{and changes} the self-resonant frequency and tunability with dc flux \cite{Trep13,Zhang15}. \textcolor{black}{The rf SQUID response is also tunable with temperature through the temperature-dependent critical current $I_c$ and inductance of the superconducting loop. }

The key enabling characteristic of the rf SQUID as a meta-atom is its nonlinearity, which has been thoroughly examined and demonstrated by the two-tone intermodulation distortion (IMD) experimental results on rf SQUID metamaterials \cite{Zhang16,Zack22}. The intrinsic nonlinearity of the Josephson effect, along with the extreme tunability of rf SQUIDs, leads to bistability \cite{Laz13,Muller19} and multistability \cite{Jung14a,Tsir14} in their  response to rf and dc driving fields. This in turn leads to complex and hysteretic behavior, including the phenomenon of transparency \cite{Zhang15}. Theory predicts that, under appropriate circumstances, driven rf SQUIDs will display strange nonchaotic attractors \cite{Zhou92} and chaos \cite{Hiz18,Shena20}. The rf SQUID metamaterials can also act as nonlinear gain media when immersed in passing electromagnetic waves \cite{Cast07,Cast08,Mack15,Zorin16,Devo16,Kis19,Aum20}, which is based on the nonlinear processes enabled by the Josephson effect that transfer energy to a signal at frequency $f_s$ from a strong pump signal at frequency $f_p$.

In early theoretical and experimental works on rf SQUID metamaterials, the rf SQUIDs were packed together side-by-side in either one \cite{Butz13,Jung13,Macha14,Besedin21} or two dimensions,\cite{Trep13}  with substantial long-range (dipole-dipole) mutual inductance of the SQUID loops due to their close lateral proximity in the plane. This coupling gives rise to remarkable collective behaviors of the metasurface, such as chimera states \cite{Laz15,Hiz16c,Hiz16s,Ban18,Hiz19Ch,Hiz20a,Hiz20b}, disorder-dominated states \cite{Laz13,Laz17,Laz18e}, and coherent modes of oscillation \cite{Trep17}.  Such states have been directly imaged by laser scanning microscopy in the superconducting state under microwave magnetic flux driving illumination \cite{JUng13c,Zhu19}. Prior work examining collections of SQUID-like entities in two dimensions, not necessarily metamaterials, include the following: superinductors made up of planar ladders of superconducting wires/loops incorporating Josephson junctions \cite{Bell12}, and Josephson transmission lines utilizing SQUID arrays to create magneto-inductive waveguides \cite{Durand92,Kuzmin19}. Another type of Josephson metamaterial recently realized utilizes a tunable plasma edge created by current-biased linear arrays of Josephson junctions embedded in a three-dimensional waveguide \cite{Trep19}. In contrast to the present work, this metamaterial interacts mainly with high frequency electric fields, rather than magnetic fields.

There have been proposals for three-dimensional versions of quantum metamaterials \cite{Zag12,Shena20}, \textcolor{black}{and previous experimental work on three dimensional superconducting metamaterials based on spiral resonators \cite{kurter_tunable_2015}. In this work, } we experimentally realized three dimensional arrays of rf SQUIDs employed as metamaterials \textcolor{black}{for the first time}. By stacking the SQUIDs vertically, we introduce \textit{positive} mutual inductive coupling for nearest neighbors, very different from the co-planar geometry and, more importantly, add the qualitatively new aspect of strong \textit{capacitive} coupling that permits high frequency currents to flow between SQUID loops. To the best of our knowledge, such coupling has not been considered in the past in any aspect of SQUID physics or technology, and can lead to dramatic new properties of coupled SQUIDs. Our three-dimensional (3D) SQUID metamaterials have flux-quantized loops mixed with non-flux-quantized loops. These latter loops are enabled by the capacitors that host displacement currents between SQUIDs.  Faraday's law is applied to all of the non-SQUID loops, in addition to the flux quantization condition in the appropriate loops.

Parasitic capacitive coupling can also occur in superconducting \textcolor{black}{digital electronics} (SCDE), based on the propagation of ps-duration single-flux quantum voltage pulses between logic circuit elements \cite{Lik91}.  
The pulses are processed by means of inductive loops, typically based on superconducting wires including Josephson junctions, essentially acting as non-resonant SQUIDs.  It is well-established that state-of-the-art SCDE suffers from an inefficient use of space on chip in many practical computing and signal processing applications. This is due to the fact that SCDE is based on magnetic flux, as opposed to the monopole electric charge utilized in CMOS electronics, and the resulting need to create and control dipole sources, such as current loops, transformers, inductors, etc. \cite{Tolypygo16}.  One way to mitigate this problem is to create three-dimensional circuit layouts in which logic and wiring layers are distributed in the third dimension, separated by ground planes \cite{Ando17}. However, this three-dimensional geometry can introduce new and unexpected coupling effects between circuits. Our inductively and capacitively coupled rf SQUID metamaterials can act as a surrogate test-bed to study coupling effects in future highly-integrated SCDE circuits.

Quantum computers utilize large arrays of qubits with controlled interactions (typically either inductive or capacitive) between many pairs of qubits \cite{Makhlin99,Wei05,Clarke08,Harris09}.  For charge and phase qubits, the nearest-neighbour interactions are enabled by capacitors, rather than inductors \cite{Berk03,Pashkin03,Averin03,McD05,Clarke08}. \textcolor{black}{Of recent interest is the design of a tunable coupler transmon that is capacitively coupled to a pair of qubits to achieve high-fidelity two-qubit gates \cite{collodo_implementation_2020,xu_high-fidelity_2020,stehlik_tunable_2021,sung_realization_2021}.} Our rf SQUID metamaterials differ in that multiple coupling capacitors are included, creating a highly integrated network of both capacitive and inductive coupling between all of the SQUIDs. We note that the effects of both capacitive and inductive coupling between rf SQUIDs has been considered as a step in deriving the quantum Hamiltonian of arrays of interacting qubits \cite{Consani20,Chitta22}. For the flux qubits, the nearest-neighbor inductive coupling can be adjusted by introducing an intermediary SQUID between the qubits to be coupled, whose properties are tuned by a local magnetic flux \cite{Fil03,Plourde04,Nisk07,Zag07}. Another approach is to have two qubits share a common wiring loop bond. This bond may have a variable kinetic inductance, or Josephson inductance, that depends on the sum of the currents flowing in the two qubit loops through that bond \cite{Majer05,Grajcar05,Grajcar06}.  Our coupling design is uniquely different in that it introduces interactions through a combination of inductive and high-frequency capacitive coupling.  The possibility exists to tune the capacitive coupling through external manipulation of the dielectric material in the capacitor.

The outline of the paper is as follows. In Section \ref{RCSJ_induct_theory}, the theory of inductively coupled rf SQUID meta-atoms is briefly reviewed.  In Section \ref{two_corner_cpl_mdl} we then introduce the strong \textit{capacitive} coupling between two SQUID loops in what we call the corner-coupled geometry and discuss its effect on the dynamics of the SQUIDs.  In Section \ref{large_system}, a sequence of larger overlapping rf SQUID structures are studied to uncover a total of three distinct classes of resonant modes from these unique metamaterial structures.  Analytical results for the resonant modes and their dispersion with dc flux are obtained in the linear response limit, and comparisons are made to full nonlinear numerical simulations.  This culminates in consideration of the N$\times$N$\times$2 metamaterial \textcolor{black} {where $2$ refers to the number of overlapping layers formed by the corner-coupled geometry} and the calculation of the number of resonant modes in such system.  In Section \ref{Expt} we then compare the results of theory to experimental data on contrasting samples: a single-layer 12$\times$12$\times$1 and an overlapping 12$\times$12$\times$2 Nb-based rf SQUID metamaterial all utilizing the same rf SQUID meta-atom.  In Section \ref{Disco}, we \textcolor{black}{discuss the generalization of our model to non-regular arrays of overlapping corner-coupled SQUIDs, } the limitations of our analysis, and opportunities for further studies of this remarkable class of superconducting metamaterials.

\section{Theory}\label{Theory}
\subsection{Model  for a system of inductively coupled SQUIDs based on the resistively and capacitively shunted junctions}\label{RCSJ_induct_theory}

First we shall review the consequences of flux quantization in a single rf SQUID loop to establish our notation and approach to setting up and solving the equations of motion for the gauge-invariant phase. The total flux $\Phi$ in the rf SQUID loop and gauge-invariant phase difference $\delta$ across the junction are related through the statement of flux quantization in a superconducting loop, which requires the order parameter to be single-valued upon going on a continuous and closed loop $C$ through the superconducting material.  Mathematically, this self-consistency condition results in \cite{Hansma73,Hansma75}, 
\begin{equation}
 2\pi n = \delta + \frac{2e}{\hbar} \Phi,
 \label{eqn:flux_quant}
 \end{equation}
where $n=0, \pm 1, \pm 2,....$ and $\Phi$ is the magnetic flux through any surface that terminates on the continuous circuit $C$.  This can be re-written as,	$\Phi = n\Phi_0 -\frac{\Phi_0}{2\pi} \delta$, where $\Phi_0=\frac{h}{2e}$ is the flux quantum.  Without loss of generality, taking $n=0$ in the SQUID loop, the expression is simplified to $\Phi = -\frac{\Phi_0}{2\pi} \delta$. We shall assume that the SQUID loop maintains quasi-static flux quantization through the microwave frequency range so that Eq. (\ref{eqn:flux_quant}) holds for both the rf and dc flux in a single galvanically-connected rf SQUID loop.  One would expect that Eq.  (\ref{eqn:flux_quant}) remains valid for all situations in which the superconducting order parameter has a well-defined phase throughout the material, which should extend to time scales as short as the order parameter relaxation time, expected to be in the ps range, except close to T$_c$ \cite{Tink96,Booth96}. 

The total flux in the loop can be expressed as \cite{Hansma73,Hansma75,RifkinrfS76,Rifkin76},
\begin{equation}
\Phi=-\Phi_\text{app} + \Phi_\text{ind}
\label{eqn:flux_relation}
\end{equation}
where $\Phi_\text{app}$ stands for the applied flux \textcolor{black}{and the minus sign is chosen to account for the diamagnetic relationship between the applied and induced fluxes}. The induced flux is $\Phi_\text{ind}=LI$ for a single SQUID with a loop inductance $L$ carrying current $I$.  Note that this relationship can be generalized to a system of many interacting SQUIDs as $\vec{\Phi}_\text{ind}=\overleftrightarrow{L}\vec{I}$, where the inductance matrix $\overleftrightarrow{L}$ contains self-inductances on its diagonal and mutual inductances between SQUIDs in the off-diagonal elements \cite{Trep17}.

The current $I$ in the junction is expressed using the resistively and capacitively shunted junction (RCSJ) model where the junction is treated as a parallel combination of three branches: an ideal Josephson junction,  a capacitor $C$ and a shunt resistor $R$ \cite{Barone82}. The total current from the three branches is thus:
$$ I=I_\text{JJ} + I_\text{R} + I_\text{C} = I_\text{c}\sin\delta +\frac{V}{R} +C\frac{dV}{dt}$$
Using the second Josephson equation, one can relate the voltage drop on the junction $V$ to the time derivative of the gauge-invariant phase as $V=\frac{\Phi_0}{2\pi} \dot{\delta}$, where the over-dot denotes time derivative.  In this case, the response current can be written as,
\begin{equation}
I=I_\text{c} \sin\delta +\frac{\Phi_0}{2\pi} \frac{\dot{\delta}}{R} +C\frac{\Phi_0}{2\pi} \ddot{\delta}
\label{eqn:Ircsj}
\end{equation}
									
The flux quantization condition, Eq. (\ref{eqn:flux_relation}) can now be written for a system of inductively-coupled identical rf SQUIDs as, \cite{Trep17}
 \begin{eqnarray}
 \vec{\Phi}_\text{dc} + \vec{\Phi}_\text{rf}\sin(\omega t)=\frac{\Phi_0}{2\pi}\vec{\delta}  +\overleftrightarrow{L}(I_\text{c} \sin\vec{\delta} +\frac{\Phi_0}{2\pi} \frac{\dot{\vec{\delta}}}{R} +C\frac{\Phi_0}{2\pi} \ddot{\vec{\delta}}), \nonumber
 \end{eqnarray}
whose normalized form reads
\begin{eqnarray} 
 \vec{\phi}_\text{dc} + \vec{\phi}_\text{rf}\sin(\Omega \tau)=\vec{\delta}  +\overleftrightarrow{\kappa}(\beta_\text{rf}\sin\vec{\delta} +\gamma \frac{d\vec{\delta}}{d\tau} + \frac{d^2\vec{\delta}}{d\tau^2}),
 \label{eqn:FluxEq}
 \end{eqnarray}
where $\vec{\Phi}_\text{dc}$ and $\vec{\Phi}_\text{rf}$ are the vectors of dc and rf magnetic flux applied to each SQUID in the array, respectively, and we assume time-harmonic rf flux at a single frequency $\omega$.  Here $\vec{\delta}$ is the array of gauge-invariant phase differences on the junctions in all of the rf SQUIDs in the array, and the over-dot $\dot{\square}$ represents derivative with respect to time.  The equation can be reduced into the dimensionless form with the following substitutions: $\phi_\text{dc,rf}=2\pi\Phi_\text{dc,rf}/\Phi_0$, $\overleftrightarrow{\kappa}=\overleftrightarrow{L}/L_\text{geo}$,  where $L_\text{geo}$ is the geometric inductance for a single SQUID in the system,  $\beta_\text{rf}=2\pi L_\text{geo}I_\text{c}/\Phi_0$, $\gamma=\sqrt{L_\text{geo}/C}/R$, $\tau=\omega_\text{geo}t= t/\sqrt{L_\text{geo}C}$, and $\Omega=\omega/\omega_\text{geo}=\omega\sqrt{L_\text{geo}C}$.  Note that we also introduce the geometric resonance $\omega_\text{geo} = 1/\sqrt{L_\text{geo}C}$, which is the resonant frequency of the rf SQUID meta-atom in the absence of the Josephson effect. 

Equation (\ref{eqn:FluxEq}) is a system of driven second-order coupled nonlinear differential equations for the gauge-invariant phase vector as a function of time, $\vec{\delta}(\tau)$, which dictates the response of the metamaterial to external electromagnetic fields.  Solving for $\vec{\delta}(\tau)$ allows one to calculate all observable properties of the system.  Prior work has explored solutions to these equations for purely inductively-coupled rf SQUID metamaterials \cite{Zhang16,Trep17,Zack22}. The following sections focus on extending this model to capacitively-coupled rf SQUIDs.  
We begin with the simplest case of a pair of corner-coupled overlapping rf SQUIDs, and then consider a sequence of larger and larger arrays of overlapping \textcolor{black}{corner-coupled} SQUIDs, eventually addressing the problem of the $N\times N\times 2$ system. \textcolor{black}{The parameters for the SQUIDs in the following calculation are given in Appendix \ref{app:design_parameters} and Table. \ref{tab:SQUIDparams}.}


\subsection{Model for two corner-coupled overlapping SQUIDs}\label{two_corner_cpl_mdl}
\begin{figure}[H]
\includegraphics[width=\columnwidth]{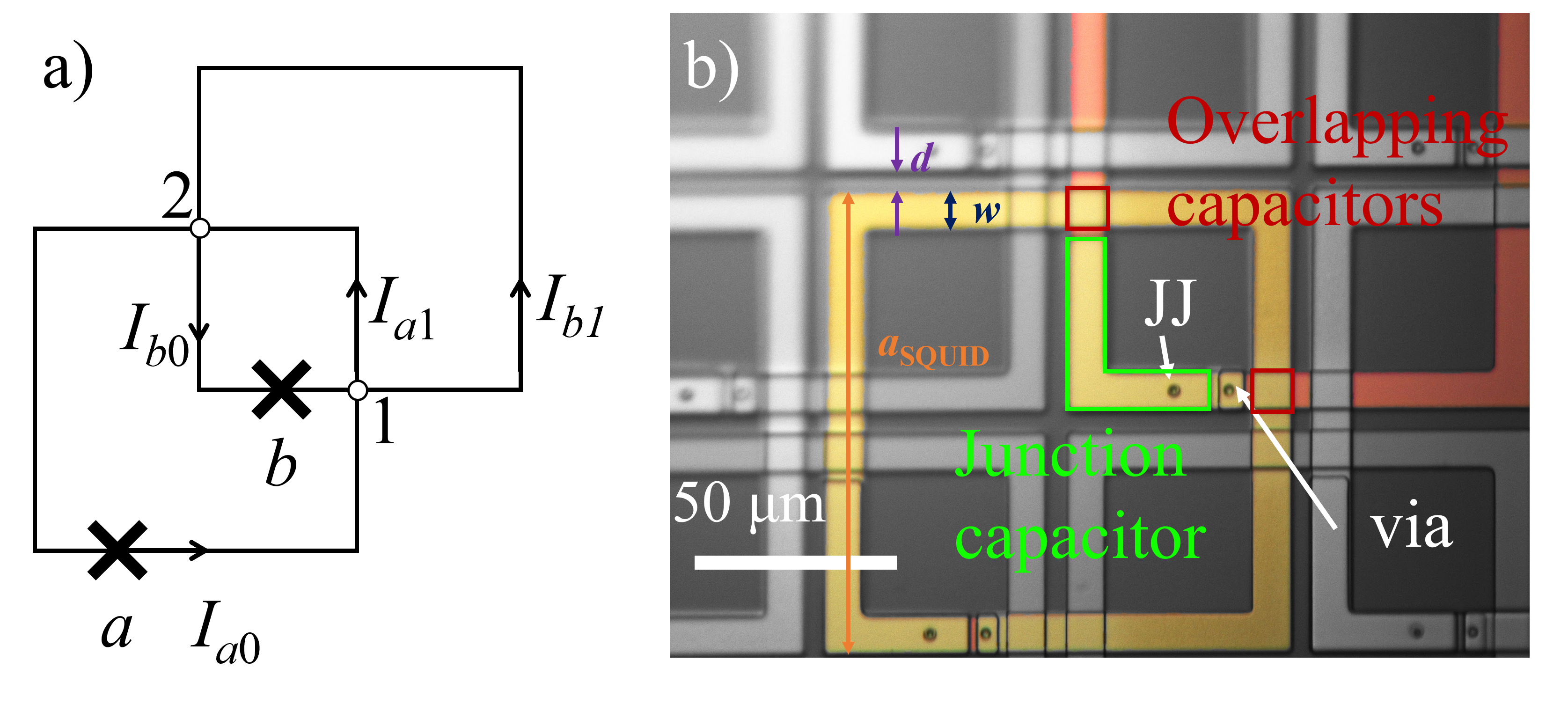}
\caption{\label{fig:corner_schem}(a) Schematic diagram of a pair of corner-coupled rf SQUIDs with overlapping wiring layers, creating capacitors labeled 1 and 2, along with the two nominally identical Josephson junctions \textcolor{black}{in loop} $a$ and \textcolor{black}{loop} $b$ represented with ``$\times$"\textcolor{black}{s}.  Junction currents $I_{a0}$ and $I_{b0}$ differ from the corresponding non-junction currents $I_{a1}$ and $I_{b1}$, in general.  (b) Micrograph of a small section of a $7\times7\times2$ metamaterial made up of corner-coupled SQUIDs (fabricated by the SNEP process), where one representative pair of the capacitively-coupled SQUIDs is highlighted. The different colors of the wiring correspond to different lithographic layers of the device.}
\end{figure}

The simplest model for overlapping SQUIDs is a pair of rf SQUID loops having wiring layers overlapping each other at the corner, and the overlapping portions are separated by a thin dielectric layer, which forms two capacitors whose capacitance $C_\text{ov}$ is comparable to that of the junction, $C$ (see Fig. \ref{fig:corner_schem} (b)).   It should be noted that these overlapping capacitors \textit{do not} include Josephson coupling between the superconducting wires, but do create a route for high-frequency displacement currents to flow between the wiring loops of neighboring SQUIDs.  The capacitors shunt the SQUID loop, breaking the uniformity of high-frequency currents in the loops, which leads to different currents in the non-junction branches $I_{a1(b1)}$ from the currents in the junction branches $I_{a0(b0)}$ as illustrated in Fig. \ref{fig:corner_schem} (a).  Note that the overlapping capacitors have no direct influence on the dc currents, which are constrained to flow \textit{only} through individual SQUID loops.

\subsubsection{Loops in the two corner-coupled SQUIDs}
Following the same flux-to-current approach in Sec. \ref{RCSJ_induct_theory}, one can write down the flux quantization conditions for the two corner-coupled SQUIDs (called $a$ and $b$) as follows:
\begin{eqnarray}
\begin{pmatrix}
    \Phi_a^\text{app}\\ \Phi_b^\text{app}
\end{pmatrix}
=\frac{\Phi_0}{2\pi}
\begin{pmatrix}
    \delta_a\\ \delta_b
\end{pmatrix}+\begin{pmatrix}
    \Phi_a^\text{ind}\\ \Phi_b^\text{ind}
\end{pmatrix} \nonumber\\
\text{with}
\begin{pmatrix}
    \Phi_a^\text{ind}\\ \Phi_b^\text{ind}
\end{pmatrix}=
\begin{pmatrix}
    L_{a,a0}& M_{a,b0}& L_{a,a1}& M_{a,b1}\\
    M_{b,a0}& L_{b,b0}& M_{b,a1}& L_{b,b1}
\end{pmatrix}
\begin{pmatrix}
    I_{a0}\\ I_{b0}\\ I_{a1}\\ I_{b1}
\end{pmatrix}
    \label{eqn:flux_relation_corner2}
\end{eqnarray}
where the induced flux is expressed as contributions from different \textit{segments} of the two SQUID loops on the second line. The elements of the first (second) row in the inductance matrix are determined as the \textit{partial inductance} between the individual segments denoted by the subscript, and the galvanically connected SQUID loop a (b). Partial inductance is a concept that generalizes the inductance of a closed loop to that of segments in the loop \cite{rosa_self_1908,paul_inductance_2009}.  Consider a segment $1$ in a closed loop $c$ that is experiencing a time-varying magnetic field due to the current in segment $2$ in a closed loop $d$.  The total flux in the closed loop \textcolor{black}{$c$} is simply $\Phi=\iint \vec B\cdot d\vec S =\oint_c \vec A \cdot d\vec l$ where the magnetic field and the vector potential are due to the current in segment $2$. We can assign the flux contribution from individual  segments as follows: $\Phi=\oint_c \vec A \cdot d\vec l=\sum_i\int_{s_i\in c}\vec A\cdot d\vec l$. The partial inductance between segments $1$ and $2$ is then defined as the ratio, $M_{1,2}=\int_1 \vec A\cdot d \vec l/I_2$. Consequently, it follows that for a closed loop $c$, $M_{c,2}=\sum_{s_i\in c}M_{s_i,2}$, and between two closed loops $c$ and $d$, $M_{c,d}=\sum_{s_j\in d} M_{c,s_j}$.

Although Eq. (\ref{eqn:flux_relation_corner2}) relates the flux with the currents, the non-junction currents $I_{a1,b1}$ are not yet expressed as functions of the gauge-invariant phase differences $\delta_{a,b}$ and their time derivatives. To obtain the equations of motion in $\vec\delta$, one needs to invoke Faraday's law on the center overlapping loop from capacitor $1$ to $2$ to junction $b$ and back to capacitor $1$ in Fig. \ref{fig:corner_schem} (a), as well as current conservation on the capacitor nodes.  This will allow us to solve for $I_{a1,b1}$ in terms of the gauge-invariant phases.

\subsubsection{Faraday's law applied to the center overlapping loop}
There is no need to consider Faraday's law in the formulation of the equations for the non-overlapping SQUIDs, since it is implicitly incorporated in the flux quantization condition Eq. (\ref{eqn:FluxEq}). In fact, applying Faraday's law to a single SQUID loop will result in the time derivative of Eq. (\ref{eqn:FluxEq}). On the other hand, its application is necessary for a superconducting loop interrupted by capacitors, such as the small center loop carrying currents $I_{b0}$ and $I_{a1}$ formed by the two overlapping corner-coupled SQUIDs in Fig. \ref{fig:corner_schem} (a). By applying Faraday's law to this center loop (denoted $\text{cen}$), one finds:
\begin{eqnarray}
&V_b - V_1 + V_2 = \dot{\Phi}_\text{cen}^\text{app}-\dot{\Phi}_\text{cen}^\text{ind}  \label{eqn:faraday_center_corner2}
\end{eqnarray}
\textcolor{black}{with} $\Phi_\text{cen}^\text{ind}=M_{\text{cen},a0}I_{a0}+L_{\text{cen},b0}I_{b0}+L_{\text{cen},a1}I_{a1}+M_{\text{cen},b1}I_{b1}$. $V_1$, $V_2$ are the voltages across the capacitors at nodes $1$ and $2$. \textcolor{black}{The sign convention of the capacitor voltages are chosen so that the positive voltage corresponds to the electric field pointing out of the plane, from the bottom \ch{Nb} wiring layer to the top \ch{Nb} wiring layer or from SQUID $b$ to SQUID $a$}. $V_b$ is the voltage across the junction of SQUID loop $b$, and the currents $I_{a0,b0,a1,b1}$ are  labeled in Fig. \ref{fig:corner_schem} (a).  Again, the partial inductances $M_{\text{cen},a0}, L_{\text{cen},b0}, L_{\text{cen},a1}, M_{\text{cen},b1}$ from each segment to the center overlapping loop (cen) are involved in \textcolor{black}{the expression for the induced flux $\Phi_\text{cen}^\text{ind}$}.

\subsubsection{Conservation of current through the overlapping capacitors}
The effect of capacitive coupling on the corner-coupled SQUIDs are understood through the conservation of currents applied to the overlapping capacitors:
\begin{subequations}
\label{eqn:current_law_corner2}
\begin{eqnarray}
    I_{a0} = I_{a1} + C_\text{ov} \dot{V}_1 \\
    I_{b0} = I_{b1} - C_\text{ov} \dot{V}_1 \\
    I_{a0} = I_{a1} - C_\text{ov} \dot{V}_2 \\
    I_{b0} = I_{b1} + C_\text{ov} \dot{V}_2,
\end{eqnarray}
\end{subequations}
where nodes 1 and 2 have identical capacitance $C_\text{ov}$ based on our design. The current conservation statements, Eqs. (\ref{eqn:current_law_corner2}), reduce to the following relations:
\begin{eqnarray}
    I_{a0} + I_{b0} = I_{a1} + I_{b1}\label{eqn:current_cons_corner2_constr}\\
    \dot{V}_1 = -\dot{V}_2\label{eqn:current_cons_corner2_v1v2}
\end{eqnarray}
The flux equation Eq. (\ref{eqn:flux_relation_corner2}) allows us to express the non-junction currents $I_{a1}$ and $I_{b1}$ in terms of the junction currents and the gauge-invariant phases, in other words $I_{a1}(I_{a0},I_{b0},\delta_a,\delta_b)$, $I_{b1}(I_{a0},I_{b0},\delta_a,\delta_b)$ as\textcolor{black}{:}
\begin{eqnarray}
I_{a1}=\text{CD}^{-1}\left[
\begin{pmatrix}
L_{b,b1} & M_{a,b1}
\end{pmatrix}
\begin{pmatrix}
    \Phi_a^\text{app}-\frac{\Phi_0}{2\pi}\delta_a\\
    -\Phi_b^\text{app}+\frac{\Phi_0}{2\pi}\delta_b
\end{pmatrix}
-\right.\nonumber\\
\left.\det
\begin{pmatrix}
    L_{a,a0}&M_{a,b1}\\
    M_{b,a0}&L_{b,b1}
\end{pmatrix}I_{a0}-\det
\begin{pmatrix}
    M_{a,b0}&M_{a,b1}\\
    L_{b,b0}&L_{b,b1}
\end{pmatrix}I_{b0}\right],\nonumber\\
I_{b1}=\text{CD}^{-1}\left[
\begin{pmatrix}
    M_{b,a1}& L_{a,a1}
\end{pmatrix}
\begin{pmatrix}
    -\Phi_a^\text{app}+\frac{\Phi_0}{2\pi}\delta_a\\
    \Phi_b^\text{app}-\frac{\Phi_0}{2\pi}\delta_b
\end{pmatrix}
-\right.\nonumber\\
\left.\det
\begin{pmatrix}
    L_{a,a1}&L_{a,a0}\\
    M_{b,a1}&M_{b,a0}
\end{pmatrix}I_{a0}-\det
\begin{pmatrix}
    L_{a,a1}&M_{a,b0}\\
    M_{b,a1}&L_{b,b0}
\end{pmatrix}I_{b0}\right]\label{eqn:Ia1b1_from_fluxeqn}.
\end{eqnarray}
The definition of $\text{CD}$ is given in Appendix \ref{app:param_def}. The current conservation statement Eq.(\ref{eqn:current_cons_corner2_constr})  now becomes a constraint on $\delta_a, \delta_b,I_{a0},I_{b0}$ after substituting $I_{a1,b1}$ from Eqs. (\ref{eqn:Ia1b1_from_fluxeqn}):

\begin{eqnarray}
    \begin{pmatrix}
        \kappa_a & \kappa_b
    \end{pmatrix}
    \begin{pmatrix}
        I_{a0} \\ I_{b0}
    \end{pmatrix}=
    \begin{pmatrix}
        L^{-1}_{\delta a} & L^{-1}_{\delta b}
    \end{pmatrix}
    \begin{pmatrix}
        \Phi^\text{app}_a-\frac{\Phi_0}{2\pi}\delta_a\\ \Phi^\text{app}_b-\frac{\Phi_0}{2\pi}\delta_b
    \end{pmatrix},\quad
    \label{eqn:eom_con_corner2}
\end{eqnarray}
where $\kappa_{a,b}$ and $L_{\delta a,\delta b}$ are defined in Appendix \ref{app:param_def}. \textcolor{black}{After substituting the expressions for $I_{a1,b1}$ in Eqs. (\ref{eqn:Ia1b1_from_fluxeqn}) into Faraday's law Eq. (\ref{eqn:faraday_center_corner2}), the time derivatives of junction voltages in Eq. (\ref{eqn:current_cons_corner2_v1v2}), $\dot{V}_1=-\dot{V}_2$, are determined as:}
 \begin{eqnarray}
 \dot{V}_1=[(1+\kappa_{\text{v}b})\dot{V}_b+\kappa_{\text{v}a}\dot{V}_a+L_{\text{I}b}\ddot{I}_{b0}+L_{\text{I}a}\ddot{I}_{a0}\nonumber\\
-\ddot{\Phi}^\text{app}_\text{cen}-\kappa_{\text{v} a}\ddot{\Phi}^\text{app}_{a}-\kappa_{\text{v} b}\ddot{\Phi}^\text{app}_{b}]/2\label{eqn:dv1_2_corner}
 \end{eqnarray}

The parameters $\kappa_{\text{v}a}, \kappa_{\text{v}b}, L_{\text{I}a}, L_{\text{I}b}$ are defined in Appendix \ref{app:param_def}. One can see that Eq. (\ref{eqn:dv1_2_corner}) \textcolor{black}{contains 4th-order derivatives of} $\delta_{a,b}$, by noting that $I_{a0,b0}$ given in the RCSJ model Eq. (\ref{eqn:Ircsj}) brings two more time derivatives into the equation. The expression for $\dot{V}_1$ Eq. (\ref{eqn:dv1_2_corner}) can then be substituted into the current laws Eq. (\ref{eqn:current_law_corner2}) to obtain solutions for the non-junction currents $I_{a1,b1}(I_{a0,b0},\ddot{I}_{a0,b0},\ddot{\delta}_{a,b})$.
\subsubsection{Equation of motion for gauge invariant phase differences}
The flux equations for the two SQUID loops can now be set up. Assuming that the applied dc and rf flux amplitudes are the same in both SQUID loops, and that the rf flux is sinusoidal at a single frequency $\omega$ with amplitude $\Phi_\text{rf}$, the flux equation, Eq. (\ref{eqn:flux_relation_corner2}), becomes:
\begin{eqnarray}
\begin{pmatrix}
    \Phi_\text{dc} +\Phi_\text{rf} \sin (\omega t)\\ \Phi_\text{dc} +\Phi_\text{rf} \sin (\omega t)
\end{pmatrix}=\frac{\Phi_0}{2\pi}
\begin{pmatrix}
    \delta_a\\ \delta_b
\end{pmatrix}+\nonumber \\
\begin{pmatrix}
    L_\text{geo}& M\\
    M& L_\text{geo}
\end{pmatrix}
\begin{pmatrix}
    I_{a0}\\I_{b0}
\end{pmatrix}+
C_\text{ov}
\begin{pmatrix}
    -L_{\delta a}\dot{V}_1\\ L_{\delta b}\dot{V}_1
\end{pmatrix},
   \label{eqn:eom_corner2}
    \end{eqnarray}
where the induced flux (last two terms on the right hand side of Eq. (\ref{eqn:eom_corner2})) is separated into two contributions: the conventional inductively-coupled SQUIDs with mutual inductance $M$, and the correction due to the overlapping capacitors. The term with the inductance matrix can be regarded as the limit without capacitive coupling, when $C_\text{ov}=0$. Consequently, the current becomes uniform inside the SQUID loops such that  $I_{a0}=I_{a1}$, $I_{b0}=I_{b1}$. The $2\times4$ inductance matrix in Eq. (\ref{eqn:flux_relation_corner2}) is then reduced to the $2\times2$ matrix above with self inductance of the SQUID loop determined as $L_\text{geo}=L_{a,a0}+L_{a,a1}=L_{b,b0}+L_{b,b1}$, and the mutual inductance between the two SQUID loops $M=M_{b,a0}+M_{b,a1}=M_{a,b0}+M_{a,b1}$, which can be positive, zero, or negative depending on the overlapping area between the two SQUID loops.  The last term in Eq. (\ref{eqn:eom_corner2}) involving $C_\text{ov}$ brings in qualitatively new phenomena in the high frequency response of coupled rf SQUIDs.
 
\subsubsection{Linear-limit solutions}\label{lin_lim_sol}
\begin{figure}[H]
\includegraphics[width=\columnwidth]{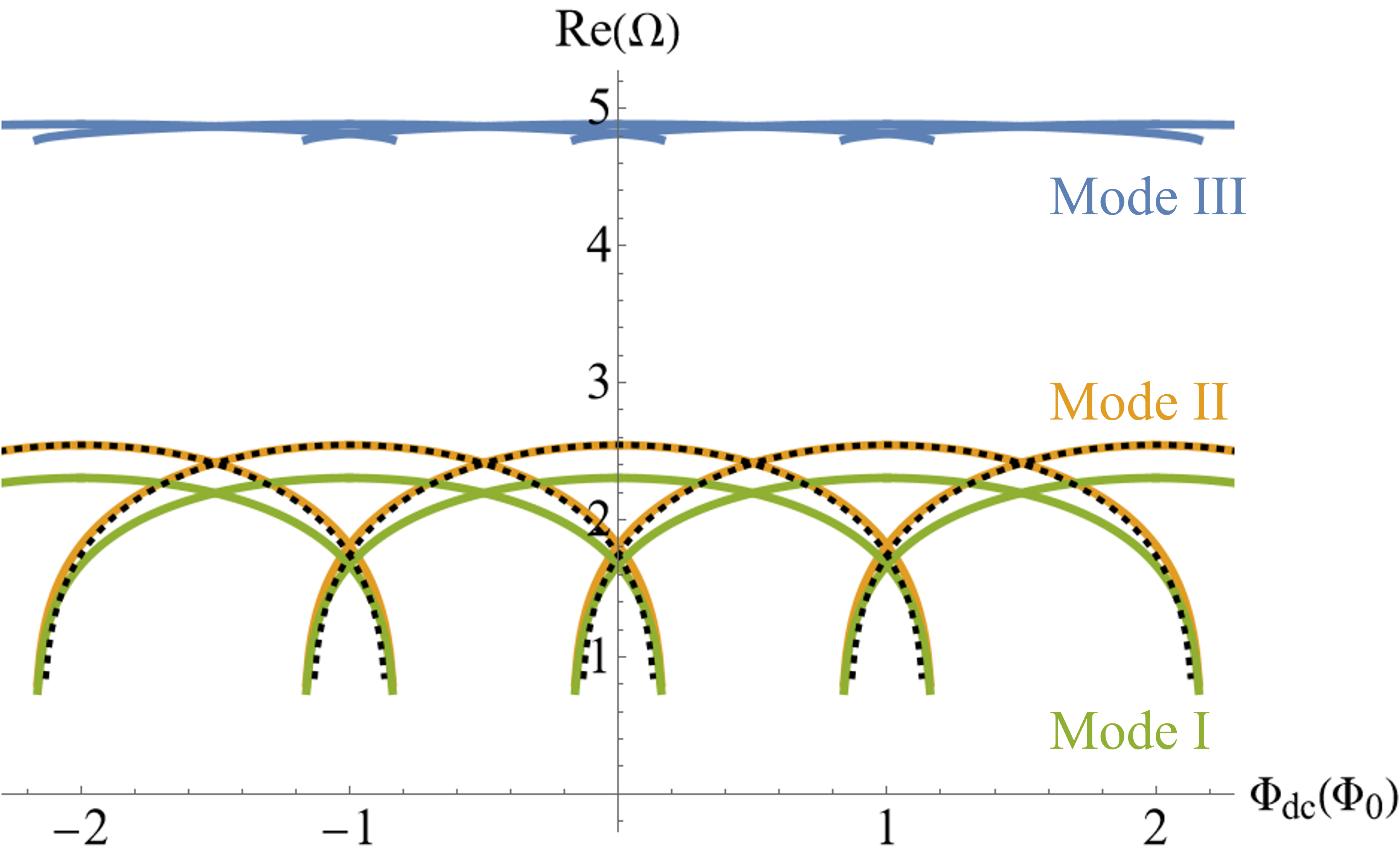}
\caption{\label{fig:lin_single_corner2}{Real part of eigenfrequency solutions $Re(\Omega)$ from the characteristic equation $\det(\overleftrightarrow\chi)=0$ for the two corner-coupled SQUIDs, as a function of dc magnetic flux $\Phi_\text{dc}$. The parameters for this calculation are given in Appendix \ref{app:design_parameters} and Table.\ref{tab:SQUIDparams}.  The solid curves with three different colors correspond to the three positive solutions to the characteristic equation, while the black dotted curve is the eigenfrequency for a single SQUID with the same parameters.  Due to the hysteretic response of the  SQUIDs ($\beta_\text{rf}>1$), multiple dc flux tuning curves overlap each other in the same range of applied dc flux. }}
\end{figure}

To analytically understand the two corner-coupled SQUID system, one can start by simplifying the equations in the low-driving-amplitude linear limit when $|\phi_\text{rf}|\ll1$. The full solution to $\delta(t)$ of any individual SQUID can be separated into its dc and rf components: $\delta=\delta_\text{rf}(t)+\delta_\text{dc}$ \cite{Zhang16}. Under a weak applied rf flux, the rf response is also vanishingly small, $|\delta_\text{rf}|\ll1$, so that any nonlinear rf response is negligible. Therefore, the solution takes the following form $\delta=\delta_\text{dc}+\delta_\text{rf}\exp(i\omega t)$, where $\omega$ is the driving frequency \cite{Zhang16}. The junction current from the RCSJ model under the weak rf flux approximation is $I=I_\text{dc}+I_\text{rf}(t)=I_c\sin (\delta_\text{dc})+I_c\cos (\delta_\text{dc}) \delta_\text{rf}(t)+i\omega \Phi_0/(2\pi R) \delta_\text{rf}(t)-\omega^2\Phi_0 C/(2\pi)\delta_\text{rf}(t)$, where the terms with second or higher order in $\delta_\text{rf}$ are dropped.  After substituting the expressions for $I_{a0}$ and $I_{b0}$ in the equation of motion Eq. (\ref{eqn:eom_corner2}), and converting to the \textcolor{black}{dimensionless} vector format, one obtains a system of  algebraic equations:
\begin{align}
\vec{\phi}_\text{dc}=\vec\delta_\text{dc}+\beta_\text{rf}\overleftrightarrow{\kappa}\sin\vec\delta_\text{dc}\nonumber\\    
\vec{\phi}_\text{rf}=(\overleftrightarrow{1}-\alpha\Omega^2\lambda_\text{cov}\overleftrightarrow{\kappa}_\text{loop})^{-1}\overleftrightarrow{\chi} \vec{\delta}_\text{rf}, \label{eqn:lin_eom_corner2} \\ 
\text{with }\overleftrightarrow{\chi}=
\overleftrightarrow{1}+\overleftrightarrow{\kappa}\beta_\text{rf}\text{diag}(\cos \vec{\delta}_\text{dc})+ i\gamma\overleftrightarrow{\kappa}\Omega-\nonumber\\
\left(\overleftrightarrow{\kappa}+\lambda_\text{cov}\overleftrightarrow{\kappa}_\text{loop}(\overleftrightarrow{\kappa}_{\delta}+\overleftrightarrow{\kappa}_\text{I}\beta_\text{rf}\text{diag}(\cos \vec{\delta}_\text{dc}))\right)\Omega^2\nonumber\\
+\lambda_\text{cov}\overleftrightarrow{\kappa}_\text{loop}\overleftrightarrow{\kappa}_\text{I} (-i\gamma \Omega^3+\Omega^4)\label{eqn:lin_eom_chi}
\end{align}
where $\overleftrightarrow{\kappa}$, $\overleftrightarrow{\kappa}_\text{loop}$, $\overleftrightarrow{\kappa}_{\delta}$, and $\overleftrightarrow{\kappa}_\text{I}$ are defined in Appendix \ref{app:param_def}.  Here $\overleftrightarrow{\kappa}$ is the $2\times2$ conventional inductive coupling matrix for the two SQUIDs without capacitive coupling, as in the middle term on the right hand side of Eq. (\ref{eqn:eom_corner2}). \textcolor{black}{Consider the case when the applied flux $\vec{\phi}_\text{rf}$ is zero, the resonance condition occurs when nontrivial solutions for $\vec\delta_\text{rf}$ exist, which requires \textcolor{black}{a non-invertible response tensor}, or $\det(\overleftrightarrow{\chi})=0$.} The real parts of the solutions to this characteristic equation in $\Omega=\omega/\omega_\text{geo}$ are the resonance frequencies, plotted in Fig. \ref{fig:lin_single_corner2} as a function of dc flux applied to the SQUIDs.  

The characteristic equation is only sixth order in $\Omega$, since the matrix $\overleftrightarrow{\kappa}_\text{I}$ in front of the  $(-i\gamma \Omega^3+\Omega^4)$ term in Eq. (\ref{eqn:lin_eom_chi}) is \textcolor{black}{non-invertible with zero determinant}. There are thus six roots from solving the sixth order equation, \textcolor{black}{which} come in pairs where the real parts are opposite to each other. Only the three positive roots are shown in Fig. \ref{fig:lin_single_corner2}. The tuning curves of eigenfrequencies are periodic in applied dc flux with a periodicity of $\Phi_0$, with each curve centered at an integer multiple of $\Phi_0$. However, each resonant solution curve extends beyond the \textcolor{black}{dc flux range of $1 \Phi_0$ }and overlaps the adjacent curves due to the hysteresis from the SQUID loop, since $\beta_\text{rf}>1$. 

\begin{figure}[H]
 \centering
\subfloat{
  \subfigimg[width=0.5\columnwidth]{ \qquad a)}{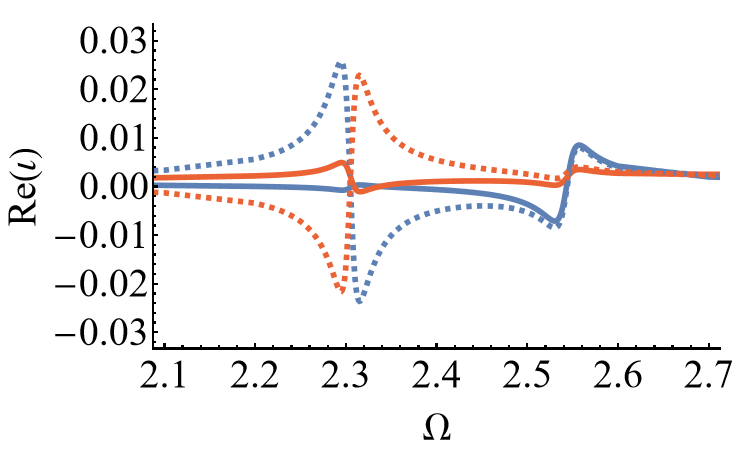}}
  \subfloat{
  \subfigimg[width=0.5\columnwidth]{\qquad c)}{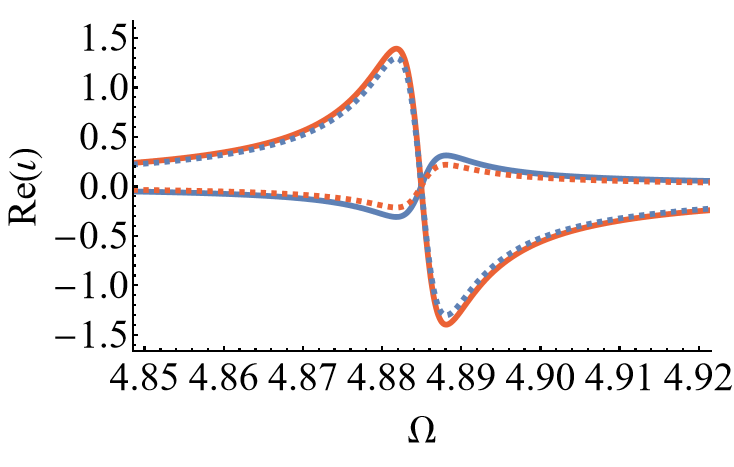}}
  \newline
\subfloat{ 
\subfigimg[width=0.5\columnwidth]{\qquad b)}{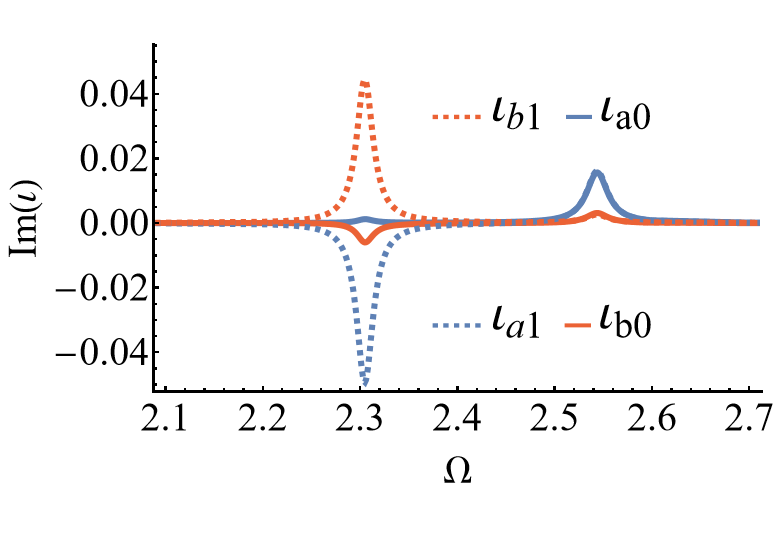}}
\subfloat{ 
\subfigimg[width=0.5\columnwidth]{\qquad d)}{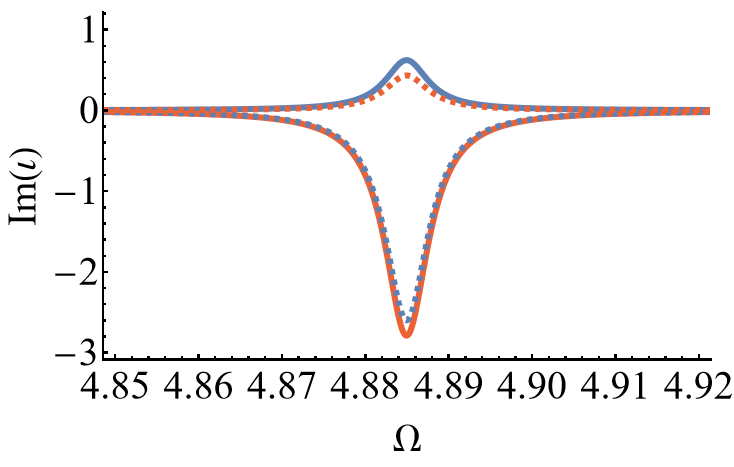}}
\caption{\label{fig:eig_currents_corner2}a), b) Real and imaginary parts of the solved dimensionless currents $\iota = 2\pi L_\text{geo}I/\Phi_0$  between $\Omega = 2.1$ and $2.7$, for the linearized case of two corner-coupled SQUIDs at zero dc flux. c), d) Real and imaginary parts of the solved currents between $\Omega = 4.8$ and $5$.  The solid curves are the solutions to the junction currents $\iota_{a0,b0}$, while the dashed curves are the non-junction currents, $\iota_{a1,b1}$. Blue curves are for the currents in loop $a$, and red for loop $b$.}
\end{figure}

The three resonances cover a much broader frequency range compared to that of a single SQUID, which is shown as a black dotted curve in Fig. \ref{fig:lin_single_corner2}. \textcolor{black}{Mode II} (the yellow curve) out of the three resonances closely follows the dc-flux tunability of a single SQUID loop resonance. To better understand the nature of the other two modes, one can examine the solutions $\vec{\delta}_\text{rf}$ to the linearized equations, Eqs. (\ref{eqn:lin_eom_corner2}) at the corresponding eigenfrequencies. In particular, the current values ($I_{a0,b0,a1,b1}$) in the SQUID loops as a function of frequency at zero dc flux are shown in Fig. \ref{fig:eig_currents_corner2}, expressed as the dimensionless currents $\iota = 2\pi L_\text{geo}I/\Phi_0$. The currents indeed undergo resonances near the three eigenfrequencies  $\Omega\sim 2.3,\ 2.55,\ 4.88$ in Fig. \ref{fig:lin_single_corner2} for zero dc flux. 

\begin{figure}[H]
    \centering
    \includegraphics[width=0.50\textwidth]{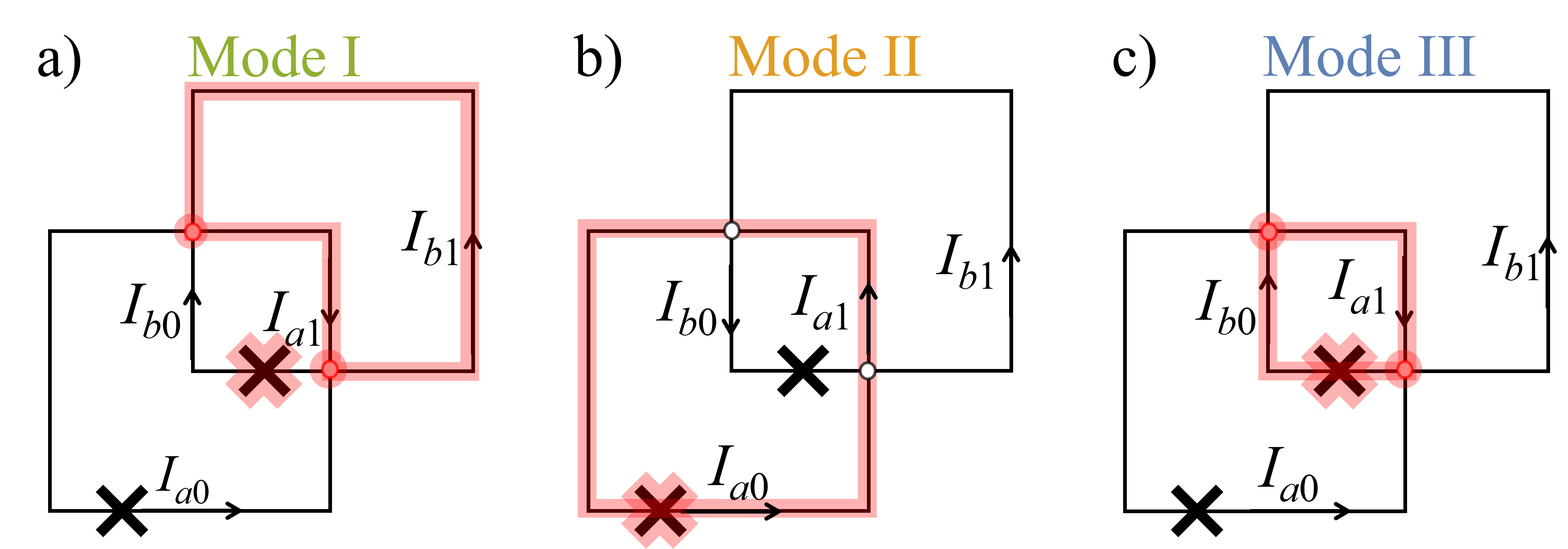}
    \caption{The dominant rf (not dc) current distribution for the three linearized eigenmodes. \textcolor{black}{The three modes are organized from left to right to match the corresponding eigenfrequencies from low to high frequency at zero dc flux in Fig. \ref{fig:lin_single_corner2}.}  The dominating junction and the ``loop'' in the mode are shaded in red. The capacitor nodes are shaded red when the rf current passes through the capacitors in the corresponding mode. }
    \label{fig:eigen_modes_corner2}
\end{figure}

After comparing the current values from the three different eigenmodes, the dominant current distribution for each mode can be summarized in the schematics in Fig. \ref{fig:eigen_modes_corner2}, where the branches with strong currents are highlighted in red. \textcolor{black}{Mode II} at $\Omega\sim2.55$ clearly stands out as the only mode where the current remains uniform inside one SQUID loop, just as in side-by-side pairs of inductively-coupled SQUIDs, which explains the match between the single SQUID eigenfrequency and the \textcolor{black}{mode II} in Fig. \ref{fig:lin_single_corner2}. 

The frequencies for the modes can also be estimated quantitatively from the current distribution in the circuit. For a single SQUID loop, the resonance can be predicted from the lumped element model, $\Omega_\text{res}=\omega_\text{res}/\omega_\text{geo}=(LC)^{-0.5}/\omega_\text{geo}=\sqrt{(L_\text{geo}^{-1}+L_\text{JJ}^{-1}) C^{-1}}/\omega_\text{geo}=\sqrt{1+\beta_\text{rf}\cos\delta}$.  This circuit model can be generalized to one SQUID loop $a$ in a large coupled system:
\begin{eqnarray}
    \Omega_\text{res}=\sqrt{(L_{a,\text{eff}}^{-1}+L_\text{JJ}^{-1}) C^{-1}}/\omega_\text{geo}  \nonumber\\
    =\sqrt{\frac{L_\text{geo}}{L_{a,\text{eff}}}+\beta_\text{rf}\cos\delta}
    \label{eqn:resonance_lin_model_Leff}
\end{eqnarray}
where the loop inductance has changed from the geometric inductance $L_\text{geo}$ to the effective inductance $L_{a,\text{eff}}(\omega)=\Phi_a^\text{ind}(\omega)/I_a(\omega)$,  accounting for the coupling from other SQUID loops in the large system. For instance, in the low power linear limit, a planar inductively coupled system has antiferromagnetic couplings among the SQUIDs. Thus,  $L_\text{eff}$ is always lower than $L
_\text{geo}$, resulting in a slightly higher $\Omega_\text{res}$ than expected for the single SQUID design parameter. This property no longer holds true in an overlapping system. In particular, for the two corner-coupled SQUIDs model, $\Phi_a^\text{ind}$ is given in Eq. (\ref{eqn:flux_relation_corner2}). The resulting $L_\text{eff}$ and $\Omega_\text{res}$ calculated for the loop $a$ and $b$ at the three resonance modes are listed in Table \ref{tab:eigen_freq_calc}. As a consequence of the nonuniform current distribution in one SQUID loop, the real part of the effective inductance can take on much wider range of values from higher than $L_\text{geo}$ to large negative values. The corresponding resonant frequencies agree with the eigenfrequencies $\Omega_\text{eig}$ solved from the linearized characteristic equation shown in Fig. \ref{fig:lin_single_corner2}. 

\begin{table}[H]
    \centering
    \begin{tabular}{|c|c|c|c|}
    \hline
        Mode &  \textcolor{black}{I}&  \textcolor{black}{II}&  \textcolor{black}{III} \\ \hline 
 $L_{a,\text{eff}}/L_\text{geo}$& $-5.35+1.09i$& $1.01+0.006i$&$0.054-0.001i$\\ \hline 
 $L_{b,\text{eff}}/L_\text{geo}$& $-5.15+2.04i$& $1.15-0.16i$&$0.054+0.001i$\\
         \hline
        $\text{Re}(\Omega_{a,\text{res}}(\delta=0))$&  $2.30$& $2.54$& $4.91$\\
         \hline
        $\text{Re}(\Omega_{b,\text{res}}(\delta=0))$&  $2.31$& $2.52$& $4.90$\\
         \hline
        $\text{Re}(\Omega_\text{eig}(\delta=0))$&  $2.31$& $2.54$& $4.89$\\
        \hline
    \end{tabular}
    \caption{Effective inductance values of SQUID loops $a$ and $b$, $L_{\text{a(b),eff}}$, in the three resonant modes of the corner-coupled rf SQUIDs.  Comparison between the real parts of the resonant frequencies $\text{Re}(\Omega_{a(b),\text{res}})$ calculated from the effective inductance for the SQUID loop $a(b)$, and the real part of their corresponding eigenfrequencies $\text{Re}(\Omega_\text{eig})$ deduced from solutions to Eqs. (\ref{eqn:lin_eom_corner2}) at zero applied dc flux.} 
    \label{tab:eigen_freq_calc}
\end{table}


\textcolor{black}{The inductance of a circuit generally scales with its size. Therefore, the longer dominant current branches in modes I \& II in Fig. \ref{fig:eigen_modes_corner2} contribute to larger effective inductance magnitudes compared to mode III, where the short segments dominate the current distribution. A small positive effective inductance, $0<L_\text{eff}\ll L_\text{geo}$, leads to a resonance frequency much higher than that of the single SQUID, $\sqrt{1+\beta_\text{rf}\cos\delta}$ . In contrast, the larger magnitudes of the effective inductance, $|L_\text{eff}|\gtrsim L_\text{geo}$, in modes I \& II in Fig. \ref{fig:eigen_modes_corner2} lead to lower resonance frequencies close to the prediction for a single SQUID.} The apparent difference in dc flux tunability between the highest frequency mode and the other modes can also be explained by the magnitude of effective inductance. The small effective inductance leads to a large $L_\text{geo}/L_\text{eff}$ that renders the dc tuning term represented by $\beta_\text{rf}\cos\delta$ less effective in Eq. (\ref{eqn:resonance_lin_model_Leff}).

\subsection{Full nonlinear numerical solutions to the two corner-coupled SQUIDs}\label{nonlin_mdl_2_corner_cpl}
Although analytical solution to the system of equations at rf flux driving levels beyond the linear limit is difficult, we can obtain the full nonlinear solution numerically. For the convenience of the numerical solver, the equations are first converted into dimensionless form as in Eq. (\ref{eqn:FluxEq}) with the additional introduction of  dimensionless currents: $\iota=2\pi L_\text{geo} I/\Phi_0$. In the general practice of numerically solving a system of differential equations,  the equations are first reformulated as a system of first order initial value problems. The equations of motion Eq. (\ref{eqn:eom_corner2}) consist of two flux equations, each a 4th-order differential equation for $\delta$. However, due to the constraint in Eq. (\ref{eqn:eom_con_corner2}) relating $\delta_{a,b}$ and $I_{a0,b0}$, there are only six degrees of freedom, two less than otherwise expected. This can be illustrated by manipulating the matrix expression in Eq. (\ref{eqn:eom_corner2}) as follows: $L_{\delta a}^{-1} \text{Row 1}+L_{\delta b}^{-1} \text{Row 2}$, which is equivalent to the constraint in Eq. (\ref{eqn:eom_con_corner2}).  Therefore, instead of solving the overdetermined system in Eq. (\ref{eqn:eom_corner2}) directly with eight variables, one should reduce the system to one equation of motion for one of the SQUIDs, along with the constraint Eq. (\ref{eqn:eom_con_corner2}), and establish the initial value problems with  six variables: $\delta_{a,b},\ \dot{\delta}_{a,b},\ \iota_{a0},\ \dot{\iota}_{a0}$, as follows.
 
\begin{subequations}
\label{eqn:eom_num_2_corner}
\begin{align}
&\frac{d\delta_a}{d\tau}=\dot{\delta}_a \\
&\frac{d\delta_b}{d\tau}=\dot{\delta}_b \\
&\frac{d\dot{\delta}_a}{d\tau}=\ddot{\delta}_a=\iota_{a0}-\beta_\text{rf}\sin \delta_a-\gamma \dot{\delta}_a \\
&\frac{d\dot{\delta}_b}{d\tau}=\ddot{\delta}_b=\iota_{b0}-\beta_\text{rf}\sin \delta_b-\gamma \dot{\delta}_b=\nonumber\\
&\kappa^{-1}_b
\begin{pmatrix}
    \kappa^{-1}_{\delta a}\\
    \kappa^{-1}_{\delta b}
\end{pmatrix}\cdot(\vec{\phi}^\text{app}-\vec{\delta})
-\frac{\kappa_a}{\kappa_b} \iota_{a0} -\beta_\text{rf}\sin \delta_b-\gamma \dot{\delta}_b\\
&\frac{d\iota_{a0}}{d\tau}=\dot{\iota}_{a0}  \\
&\frac{d\dot{\iota}_{a0}}{d\tau}=\ddot{\iota}_{a0}=\frac{1}{\lambda_\text{cov}\kappa_{\delta a}(\kappa_a \kappa_{\text{I}b}-\kappa_b \kappa_{\text{I}a})}\nonumber\\
&\left[ \begin{pmatrix}
\kappa_b-\kappa_1/\kappa_{\delta a}\\
-\kappa_1/\kappa_{\delta b}
\end{pmatrix}
\cdot (\vec{\phi}^\text{app} -\vec{\delta}) \right.\nonumber\\
&-(\kappa_b-\kappa_a \kappa_1)\iota_{a0}\nonumber\\
&\left. +\lambda_\text{cov}\kappa_{\delta a} (
\begin{pmatrix}
    \kappa_{\text{v}a}\kappa_b-\kappa_{\text{I}b}/\kappa_{\delta a}\\
    \kappa_b+\kappa_{\text{v}b}\kappa_b-\kappa_{\text{I}b}/\kappa_{\delta b}
\end{pmatrix}\cdot \ddot{\vec{\delta}}-\alpha\ddot\phi^\text{app}) \right]  
\end{align} 
\end{subequations}
where $\phi^\text{app}=\phi_\text{dc}+\phi_\text{rf}\sin (\Omega\tau)$  is the dimensionless applied flux.
The second time derivatives in $\delta$ in Eqs. (\ref{eqn:eom_num_2_corner} c, d) are related to the currents using the RCSJ model, Eq. (\ref{eqn:Ircsj}). The constraint Eq. (\ref{eqn:eom_con_corner2}) is invoked to express $\iota_{b0}$ in Eq. (\ref{eqn:eom_num_2_corner} d). The second time derivative of current ($\ddot{\iota}_{a0}$) in Eq. (\ref{eqn:eom_num_2_corner} f) is obtained from the equation of motion Eq. (\ref{eqn:eom_corner2}). All other parameters are defined in Appendix \ref{app:param_def}.  The resulting system of initial value problems was solved with the LSODA function from SciPy based on the FORTRAN library ODEPACK.


To compare with experimental results, one needs to convert the solutions for $\vec\delta(t)$ into measurable quantities.  Here, the power dissipated by the resistive channel of the junction, and the resulting change in transmission magnitude through the metamaterial are calculated as \cite{Zhang15},  
\begin{eqnarray}
P_\text{dissipation}=\sum_i\frac{V_i^2}{R}=\sum_i\left(\frac{\Phi_0 \dot{\delta_i}}{2\pi}\right)^2/R \label{eqn:dissp_pow}\\
|S_{21}|=10 \log_{10} (1-P_\text{dissipation}/P_{\text{incident}}),
\label{eqn:s21_from_dissp_pow}
\end{eqnarray}
where the sum is over Josephson junctions in the metamaterial (here $i=a,b$ and $V_i$ is the voltage drop on junction $i$), $P_\text{incident}$ is the total rf power incident on the metamaterial (which provides the rf flux bias to the SQUIDs), and $S_{21}$ is the transmission coefficient through the metamaterial at frequency $f$.  $P_\text{incident}$ is related to the applied rf flux $\Phi_\text{rf}$ as follows. For the experimental setup in Fig. \ref{fig:exp_schematics}, the sample lies in the center of a rectangular waveguide and is perpendicular to the rf magnetic field of the propagating TE$_{10}$ mode.  The rf magnetic field, and thus the rf flux $\Phi_\text{rf}$ at the location of the SQUIDs, can be calculated from the incident power $P_\text{incident}$, the waveguide dimensions, and the frequency  \cite{Trep13}.  The total dissipated power is summed over the individual contribution for each SQUID $i$. This calculation assumes that the only lossy element in the setup is the normal charge carriers tunneling in the junction, represented by the parameter $R$. 

\begin{figure}[H]
 \centering
\includegraphics[width=\columnwidth]{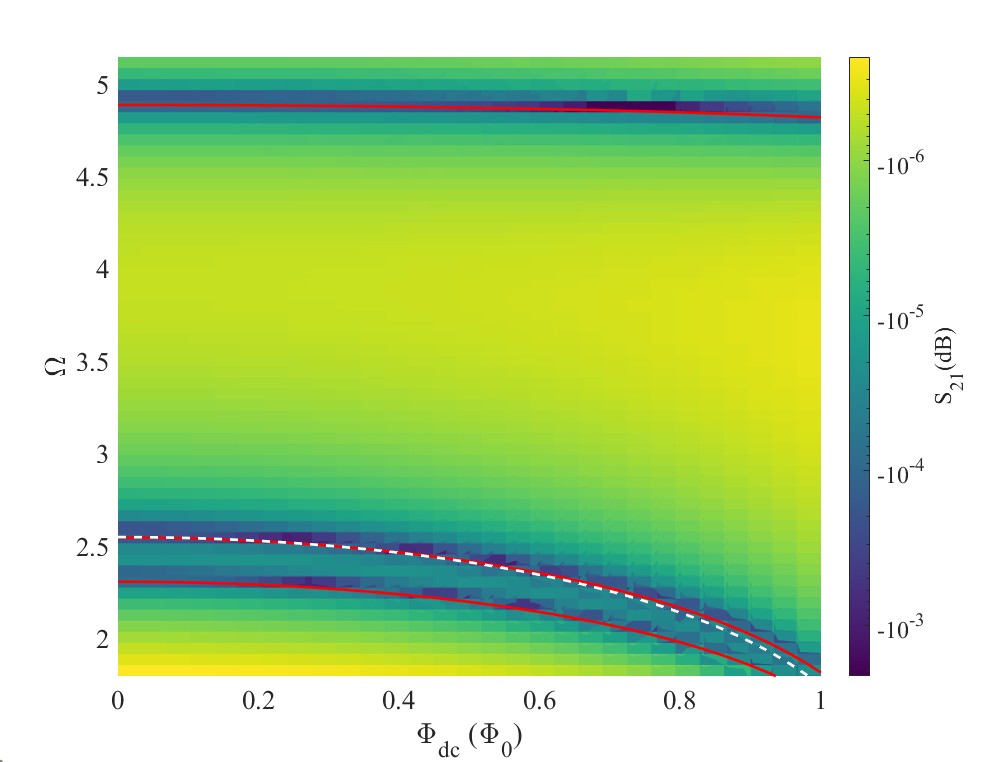}
\caption{\label{fig:nlin_sol_2_corner}{Nonlinear solutions to the two corner-coupled SQUIDs at $\Phi_\text{rf}\sim10^{-3}\Phi_0$. The parameters for this calculation are given in Appendix \ref{app:design_parameters} and Table.\ref{tab:SQUIDparams}. The quantity plotted is $|S_{21}| (\text{dB})$ on a logarithmic color scale, as a function of dimensionless frequency $\Omega = \omega/\omega_\text{geo}$ and applied dc magnetic flux in units of the flux quantum $\Phi_0$. The white dashed curve corresponds to the eigenfrequency for a single SQUID with the same parameters, and the red curves are the eigenfrequencies from the linear limit solutions for the two corner-coupled SQUIDs as in Sec.\ref{lin_lim_sol}.}}
\end{figure}

The resulting transmission at low applied rf flux amplitude ($\Phi_\text{rf}\sim10^{-3}\Phi_0$), near the linear limit, as a function of dimensionless driving frequency $\Omega$ and dc flux is plotted in Fig. \ref{fig:nlin_sol_2_corner}. The dark bands represent the resonances in the rf SQUID system, where the amplitudes of the rf currents in the SQUID loops, and thus the dissipated powers in the junctions, are maximized. Unlike the case for a pair of side-by-side rf SQUIDs, which have two resonant modes, there are now three distinct modes tuned by the applied dc magnetic flux.  The red lines in Fig. \ref{fig:nlin_sol_2_corner} show the dispersion of the linearized solutions from Eq. (\ref{eqn:lin_eom_corner2}), and show good agreement with the solutions to the full nonlinear equations, Eqs. (\ref{eqn:eom_num_2_corner}), in the weak-driving limit.

\subsection{Nonlinear Properties of the Corner-Coupled SQUIDs}\label{two_corner_cpl_NLD}

\begin{figure}[H]
    \centering
    \includegraphics[width=1\columnwidth]{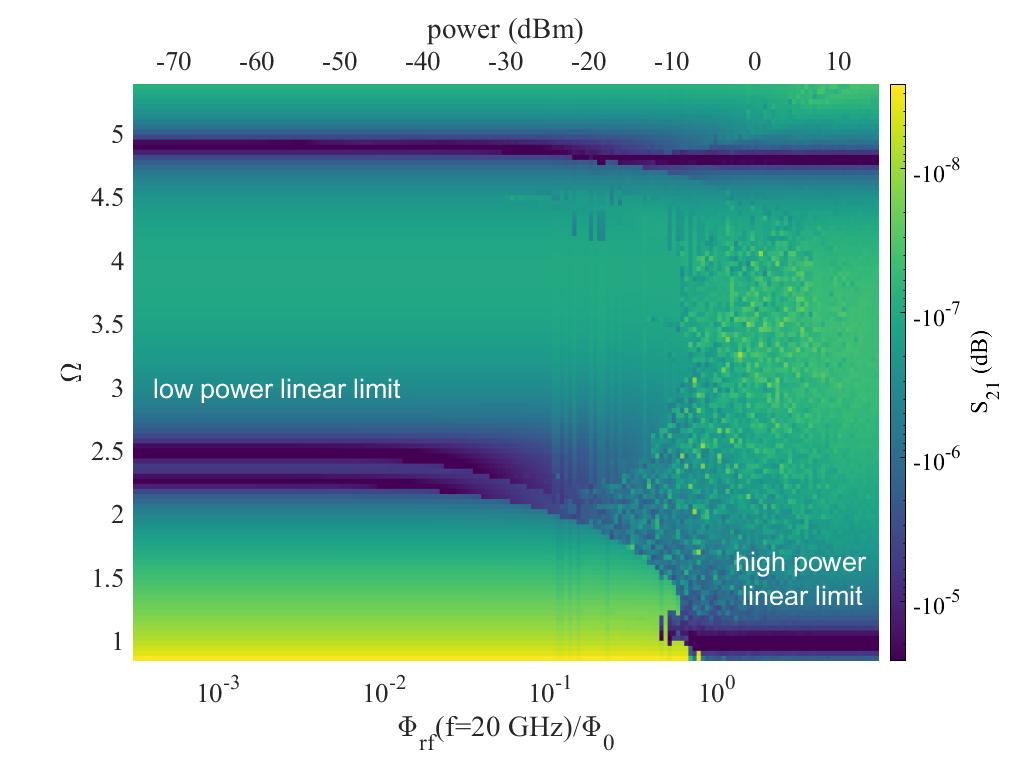}
    \caption{Nonlinear solutions to the two corner-coupled SQUIDs for $\Phi_\text{dc}=0$.  The quantity plotted is $|S_{21}| (dB)$ on a logarithmic color scale, as a function of  applied rf magnetic flux amplitude in units of the flux quantum $\Phi_0$ and dimensionless frequency $\Omega = \omega/\omega_\text{geo}$.  The linear limit discussed in Sec. \ref{lin_lim_sol} is reproduced at low applied power on the left of the plot. Upon increasing $\Phi_\text{rf}$ above $\Phi_0$, the high-power linear limit is achieved where the SQUID loop resonance is suppressed to the geometric frequency ($\Omega = 1$).}
    \label{fig:nonline_sol_2_corner_rf}
\end{figure}

Here we examine the evolution of the three modes as the amplitude of the rf driving flux is increased.  Figure \ref{fig:nonline_sol_2_corner_rf} shows the evolution of the resonant modes from the linear limit $\Phi_\text{rf}/\Phi_0 \sim 10^{-3}$ at zero dc flux to higher rf flux amplitudes.  All three modes show a suppression of their resonant frequencies in a manner similar to that observed for single rf SQUIDs \cite{Trep13}.  Note that the two lower frequency modes show substantial tuning with rf flux amplitude, but the high-frequency mode is only weakly affected \textcolor{black}{due to the small effective inductance of this mode leading to a resonance} which is less sensitive to the applied magnetic flux, \textcolor{black}{according to Eq.(\ref{eqn:resonance_lin_model_Leff})}. Also note that all three modes achieve linear response again at high driving amplitudes, in the sense that the resonant frequencies are independent of rf flux amplitude when $\Phi_\text{rf} \gtrsim \Phi_0$.  This behavior was observed before \cite{Trep13} and can be attributed to the term \textcolor{black}{linear in $\vec{\delta}$ } in Eq.  \ref{eqn:FluxEq}.  At large rf driving amplitudes, the $\sin{\vec{\delta}}$ term \textcolor{black}{is bounded between $\pm 1$, much smaller than the} leading-order $\vec{\delta}$ term, which reduces the  equation of motion to the form of a harmonic oscillator.  Further examination of the nonlinear properties of these hysteretic SQUID metamaterials will the subject of future work.

\subsection{Model for larger overlapping SQUID systems}\label{large_system}
\begin{figure}[H]
    \centering
    \includegraphics[width=1\linewidth]{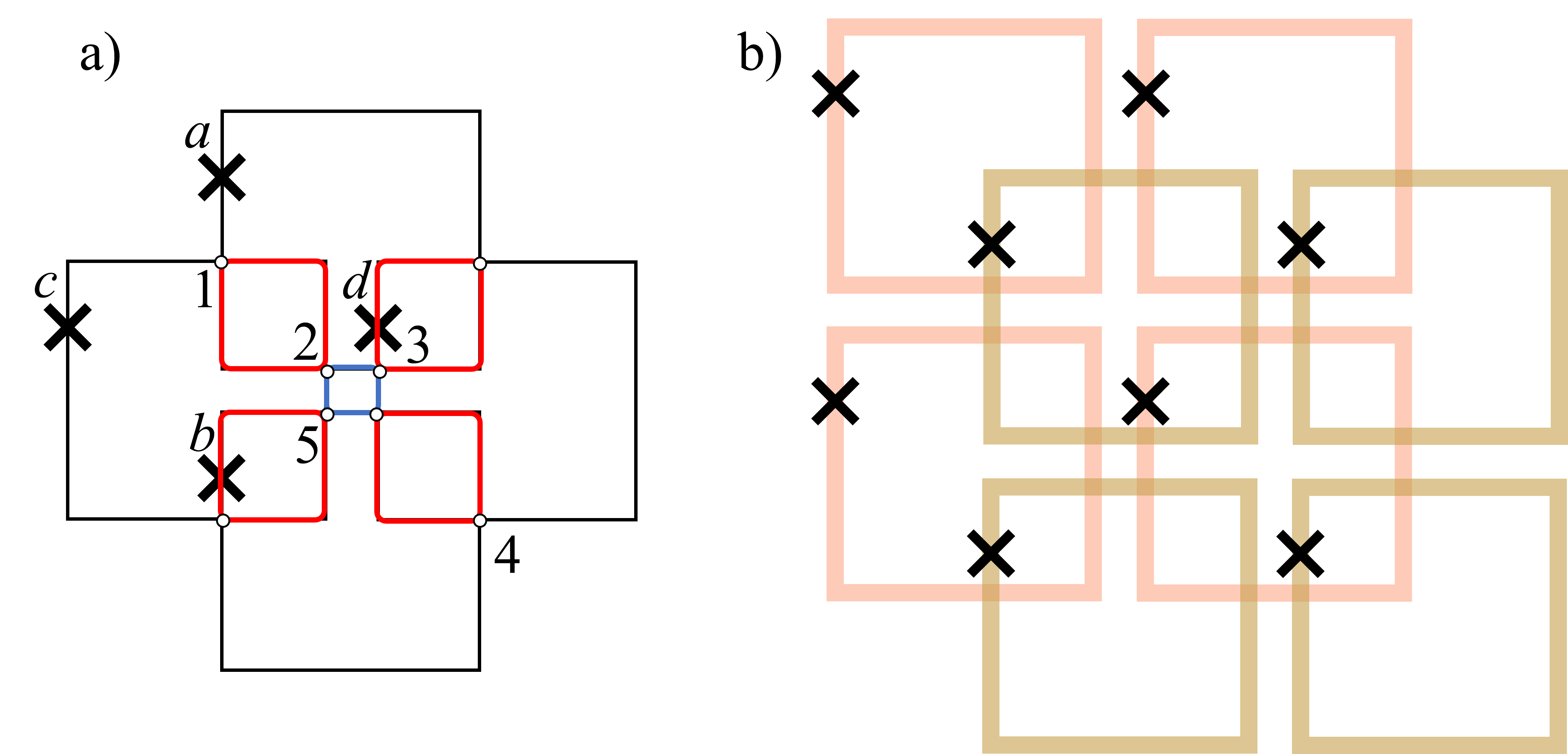}
    \caption{Schematics for larger overlapping SQUID systems.  \textcolor{black}{a) The four corner-coupled SQUIDs constitute the minimal system that contains all three different types of loops involved in the dynamics of the gauge-invariant phases. The three types of circuits are galvanically-connected SQUID loops (black lines), the partial loops (highlighted in red ), and the extra-SQUID loops (highlighted in blue).} b) represents a typical square array of $N\times N\times 2$ SQUIDs (here with $N=2$)).  \textcolor{black}{The SQUIDs from the two different layers are color coded to illustrate the geometry of this design.  Overlap capacitors exist wherever lines of different color cross.}  An array with $N=12$ has been characterized experimentally in this work.}
    \label{fig:large_SQUID_schem}
\end{figure}

Building upon the formalism developed for the two corner-coupled SQUIDs in sections \ref{two_corner_cpl_mdl}-\ref{two_corner_cpl_NLD}, we now consider the larger systems of overlapping SQUIDs as exemplified in Fig. \ref{fig:large_SQUID_schem}. One major difference in the larger systems compared to two corner-coupled SQUIDs is the introduction of a new kind of circuit loop enclosing an area outside any galvanically connected SQUID loop. To better distinguish the different loop circuits in the large systems, and to streamline the discussion, some common vocabulary should be established. Highlighted in blue in Fig. \ref{fig:large_SQUID_schem} (a) is the new loop, named the ``extra-SQUID'' loop, while the loop formed at the corners of two overlapping SQUIDs, as studied in the two-SQUID case, is referred to as a ``partial loop", colored red in Fig. \ref{fig:large_SQUID_schem} (a). Together with the conventional galvanically-connected SQUID loops, these three types of loop circuits dictate the dynamics of the gauge-invariant phases through either Faraday's law (e.g. Eq. (\ref{eqn:faraday_center_corner2})), or the flux quantization condition in the SQUID loop (e.g. Eq. (\ref{eqn:FluxEq})).

Another challenge in modelling the larger system is the fact that each SQUID loop is broken into many segments by the additional capacitive nodes formed by the overlap with wiring of several neighboring SQUIDs.  For instance, each SQUID in Fig. \ref{fig:large_SQUID_schem} (a) has $4$ capacitor nodes and thus $4$ segments. There are in total $16$ segments, each with a different current. Attempting to follow a similar treatment to that used for the two corner-coupled SQUIDs in Sec.\ref{two_corner_cpl_mdl}, one would need to express the $12$ non-junction currents in terms of $4$ junction currents, $4$ gauge-invariant phase differences and their time derivatives. The number of unknown currents can be further reduced by invoking current conservation laws on the SQUID loops requiring zero net current into any SQUID. The constraints need to be applied to all SQUID loops but one, since the net current into the last SQUID is equal to the net current leaving the rest of the SQUID loops which is already zero due to the current conservation laws. The number of independent non-junction currents is thus $12-(4-1)=9$, which still could not be completely determined from $4$ flux quantization conditions (Eq. (\ref{eqn:FluxEq})) from the $4$ SQUID loops. We should note that applying Faraday's laws to the $5$ non-SQUID loops only relates the time derivatives of the currents, not the currents themselves. 

To  circumvent this problem, the model should be reformulated in terms of the voltages across the capacitor nodes as the variables instead of the junction currents \textcolor{black}{as commonly done in the Lagrangian formalism for a magnetic circuit}. All the junction currents are expressed using the RCSJ model in Eq. (\ref{eqn:Ircsj}) in terms of the set of $\vec{\delta}$, and their time derivatives. The non-junction currents are obtained from current conservation laws on the capacitor nodes (e.g. Eq. \ref{eqn:current_law_corner2}). For the four corner-coupled SQUIDs in Fig. \ref{fig:large_SQUID_schem} (a), there are in total 8 capacitor nodes. Due to the current conservation laws mentioned above, the number of independent voltages is reduced to $5$, corresponding to the number of non-SQUID loops. This system can then be set up and solved analytically as demonstrated in Sec. \ref{four_corner_cpl_mdl}. This voltage formalism turns out to be a more general approach for modeling overlapping  SQUIDs compared to the current formalism followed in Sec. \ref{two_corner_cpl_mdl} for the two corner-coupled SQUIDs. Appendix \ref{app:volt_form_2_corner_cpl} illustrates the application of the voltage formalism to the two corner-coupled SQUIDs, and arrives at the same eigenfrequency solutions as found in Sec. \ref{two_corner_cpl_mdl}.

\subsubsection{Four corner-coupled overlapping SQUIDs}\label{four_corner_cpl_mdl}
With reference to the system shown in Fig. \ref{fig:large_SQUID_schem} (a), the dynamics of the system is now described by the $4$ gauge-invariant phase differences $\delta_{a,b,c,d}$ and their time derivatives, as well as $5$ independent capacitor nodal voltages $V_{1,2,3,4,5}$ and their time derivatives. The voltages across the rest of the capacitor nodes can be expressed in terms of $V_{1,2,3,4,5}$ through the current conservation laws inside each continuous SQUID loop. Applying Faraday's law to the non-SQUID loops and flux quantization conditions to the SQUID loops, we obtain the following system of equations of motion.
\begin{align}
    \Phi_a^\text{app}=\Phi_0\delta_a/(2\pi)+\Phi_a^\text{ind}(\vec I,\dot{\vec V})\nonumber\\
    \Phi_b^\text{app}=\Phi_0\delta_b/(2\pi)+\Phi_b^\text{ind}(\vec I,\dot{\vec V})\nonumber\\
    \Phi_c^\text{app}=\Phi_0\delta_c/(2\pi)+\Phi_c^\text{ind}(\vec I,\dot{\vec V})\nonumber\\
    \Phi_d^\text{app}=\Phi_0\delta_d/(2\pi)+\Phi_d^\text{ind}(\vec I,\dot{\vec V})\nonumber\\
    \dot{\Phi}_{ac}^\text{app}=-V_1+V_2+\dot{\Phi}_{ac}^\text{ind}(\dot{\vec I},\ddot{\vec V})\nonumber\\
    \dot{\Phi}_{ad}^\text{app}=V_d-V_3-(V_1+V_2+V_3)+\dot{\Phi}_{ad}^\text{ind}(\dot{\vec I},\ddot{\vec V})\nonumber\\
    \dot{\Phi}_{bc}^\text{app}=V_b-(V_1+V_2+V_5)-V_5+\dot{\Phi}_{bc}^\text{ind}(\dot{\vec I},\ddot{\vec V})\nonumber\\
    \dot{\Phi}_{bd}^\text{app}=(V_1+V_2-V_4)-V_4+\dot{\Phi}_{bd}^\text{ind}(\dot{\vec I},\ddot{\vec V})\nonumber\\
    \dot{\Phi}_{abcd}^\text{app}=V_5-V_2-(V_1+V_2-V_4)+V_3+\dot{\Phi}_{abcd}^\text{ind}(\dot{\vec I},\ddot{\vec V})\label{eqn:4_cnr_cpl_eom}
\end{align}
where $\vec I =(I_a,I_b,I_c,I_d)$ and $\vec V=(V_1,V_2,V_3,V_4,V_5)$. The top $4$ rows are from the flux quantization conditions on the $4$ SQUID loops, and the bottom $5$ rows from application of Faraday's law to the non-SQUID loops. The non-SQUID loops are labeled by the SQUID loops involved in forming their circuit. For example, the top left partial loop in Fig. \ref{fig:large_SQUID_schem} (a) is labelled as $ac$. The voltages in the parentheses, e.g. $(V_1+V_2+V_3)$ in the sixth row, are dependent nodal voltages obtained from applying current conservation laws to the continuous SQUID loops. The induced flux $\Phi_\text{xx}^\text{ind}$ is calculated from the partial inductances from the individual branches just as in Sec.\ref{two_corner_cpl_mdl}, which is a function of junction currents $\vec I$ and node voltages $\vec V$. 

\begin{figure}[H]
    \centering
    \includegraphics[width=0.7\linewidth]{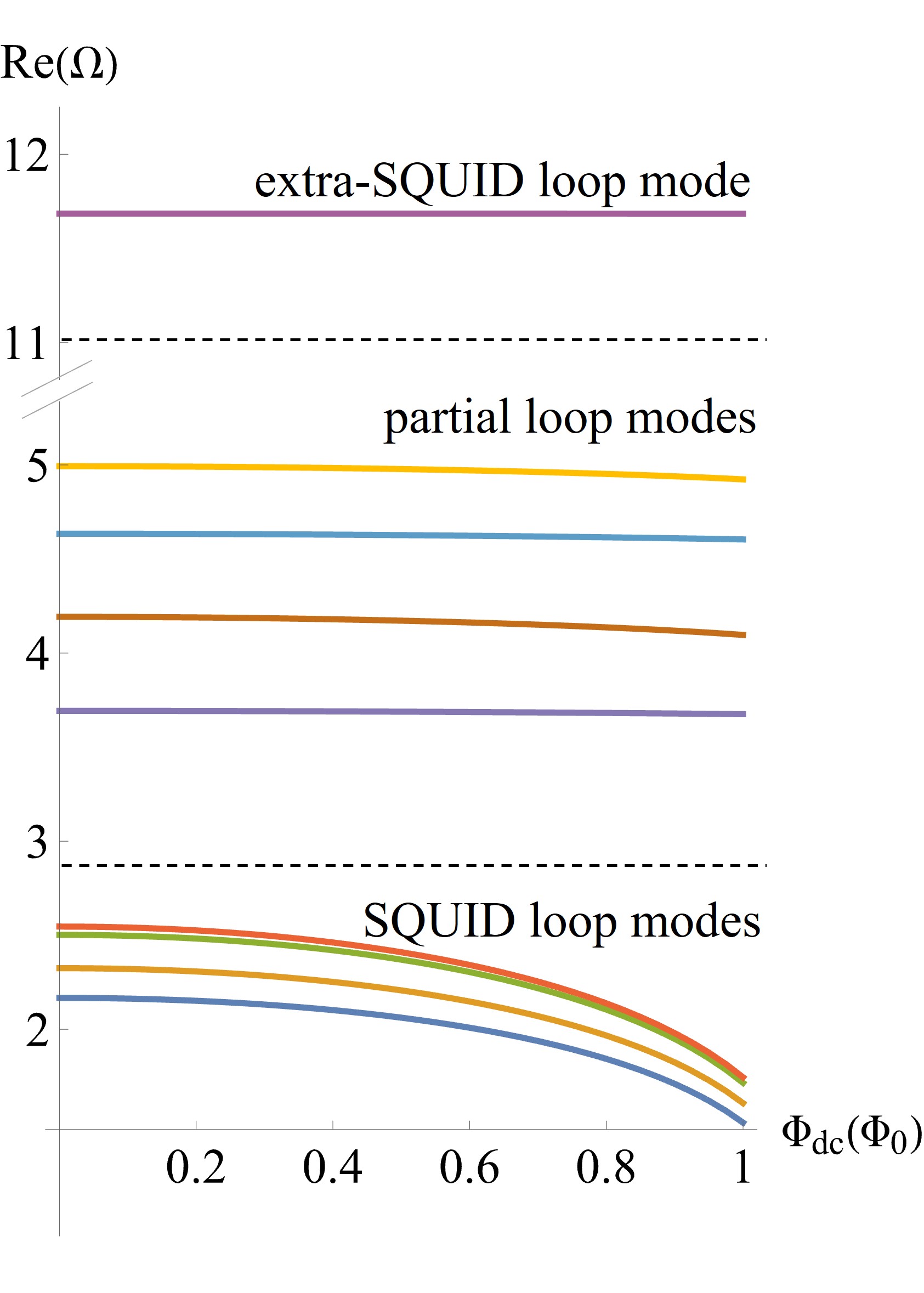}
    \caption{Real part of eigenfrequency solutions $\text{Re}(\Omega)$ from the characteristic equation $\det(\overleftrightarrow\chi)=0$ in the linear limit for the four corner-coupled SQUIDs shown in Fig. \ref{fig:large_SQUID_schem} (a), as a function of dc magnetic flux $\Phi_\text{dc}$ in units of $\Phi_0$. In this case $9$ distinct resonance modes can be resolved. The two black horizontal dashed lines delineate the three types of resonant modes.}
    \label{fig:four_corner_cpl_eig}
\end{figure}

The low rf flux amplitude linear limit solution can then be obtained by expressing  $I_{a,b,c,d}$ in terms of $\delta_{a,b,c,d}$ and their time derivatives using the RCSJ model in Eq. (\ref{eqn:Ircsj}), and replacing time derivatives with $i\Omega$ under Fourier transform. The resulting system has a size $9\times9$ with the $9$ variables: $\delta_{a,b,c,d}$ and $V_{1,2,3,4,5}$. The eigenfrequencies for this linear system can be calculated and are plotted in  Fig.  \ref{fig:four_corner_cpl_eig} as a function of dc flux. As discussed in section \ref{lin_lim_sol}, and illustrated in Figs. \ref{fig:lin_single_corner2} and  \ref{fig:eigen_modes_corner2}, the high frequency modes are dominated by the smaller loops, which contribute to smaller effective inductances, explaining their higher resonance frequency and weaker tunability, \textcolor{black}{according to } Eq. (\ref{eqn:resonance_lin_model_Leff}). The same loop-resonance correspondence can be established for the system of four corner-coupled SQUIDs containing $4$ SQUID loops, $4$ partial loops and $1$ extra-SQUID loop, as labeled in Fig. \ref{fig:four_corner_cpl_eig}.  As expected, the much smaller extra-SQUID loop brings about a very high resonance at $\Omega \approx 12$ in Fig. \ref{fig:four_corner_cpl_eig}.

\begin{figure}[H]
    \centering
    \includegraphics[width=0.7\linewidth]{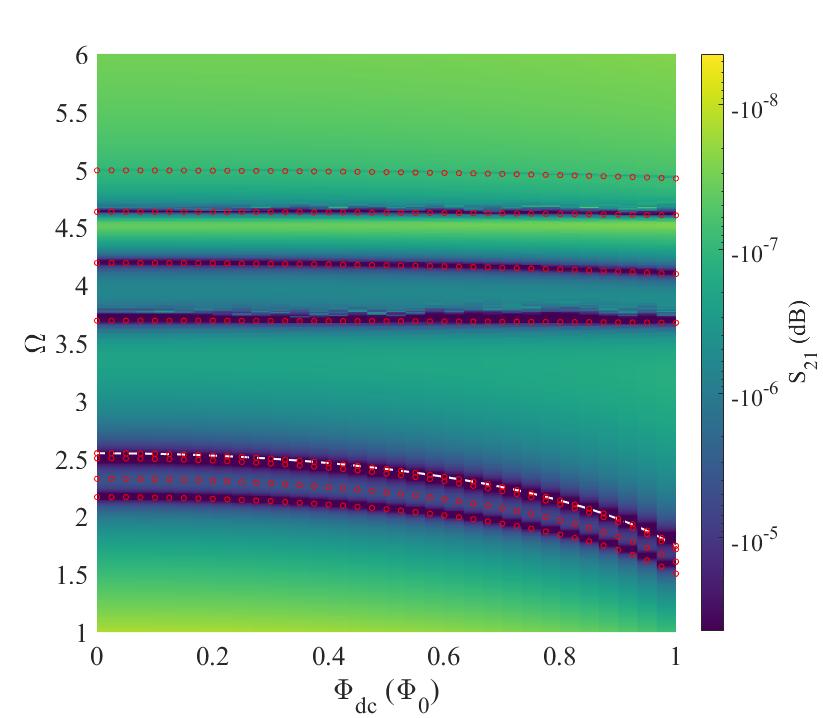}
    \caption{Nonlinear solutions to the four corner-coupled SQUIDs shown in Fig. \ref{fig:large_SQUID_schem}(a) at $\Phi_\text{rf}\sim10^{-3}\Phi_0$. The quantity plotted is $|S_{21}| (\text{dB})$ on a logarithmic color scale, as a function of dimensionless frequency $\Omega = \omega/\omega_\text{geo}$ and applied dc magnetic flux in units of $\Phi_0$. The red dotted curves are the eigenfrequencies from the linear limit solution, and are closely matching to the resonances from the numerical solution. The dashed white curve is the eigenfrequency of the single SQUID with the same design parameter. The calculation cuts off at $\Omega = 6$  below the extra-SQUID loop modes.}
    \label{fig:4_corner_cpl_nlin}
\end{figure}

Similar to the two corner-coupled SQUIDs discussed in  Sec.\ref{nonlin_mdl_2_corner_cpl}, the full nonlinear solution is obtained numerically. The details for setting up the numerical solver can be found in  Appendix \ref{app:general_volt_form_ovlp_SQUIDs}. Fig. \ref{fig:4_corner_cpl_nlin} shows the resulting transmission $|S_{21}|$ as a function of driving frequency $\Omega$ and applied dc flux $\Phi_\text{dc}$ in the low rf flux amplitude linear limit, $\Phi_\text{rf}\sim10^{-3}\Phi_0$. The \textcolor{black}{transmission calculated from the }numerical solution has good agreement with the \textcolor{black}{eigenfrequencies from the }linear limit solutions, as illustrated \textcolor{black}{by} the coincidence of the dark (absorbing) features with the red dotted curves. A separate calculation is performed around $\Omega =12$ which resolves the extra-SQUID loop modes that can not be captured in the frequency range in Fig.\ref{fig:4_corner_cpl_nlin}.

\subsubsection{\texorpdfstring{$2\times 2\times 2$}{2by2by2} corner-coupled overlapping SQUIDs} \label{2by2by2_mdl}
A full numerical solution, or even an analytical solution in the linear limit, is very computationally expensive to obtain for the large $N\times N\times2$ system studied experimentally. However, the case of $N=2$ (Fig. \ref{fig:large_SQUID_schem} (b)) can be tackled easily and should illustrate the general properties of the model. For the $2\times 2\times2$ system, there are $8$ SQUID loops, $9$ partial loops, and $2$ extra-SQUID loops.  In the absence of any symmetries, we would thus expect a total of $19$ resonant modes, comprised of $8$ lower frequency modes near the single SQUID resonance, $9$ partial loop modes at about twice the single-SQUID resonance frequency, and $2$ extra-SQUID loop modes near $\Omega=12$.

\begin{figure}[H]
    \centering
    \includegraphics[width=0.9\linewidth]{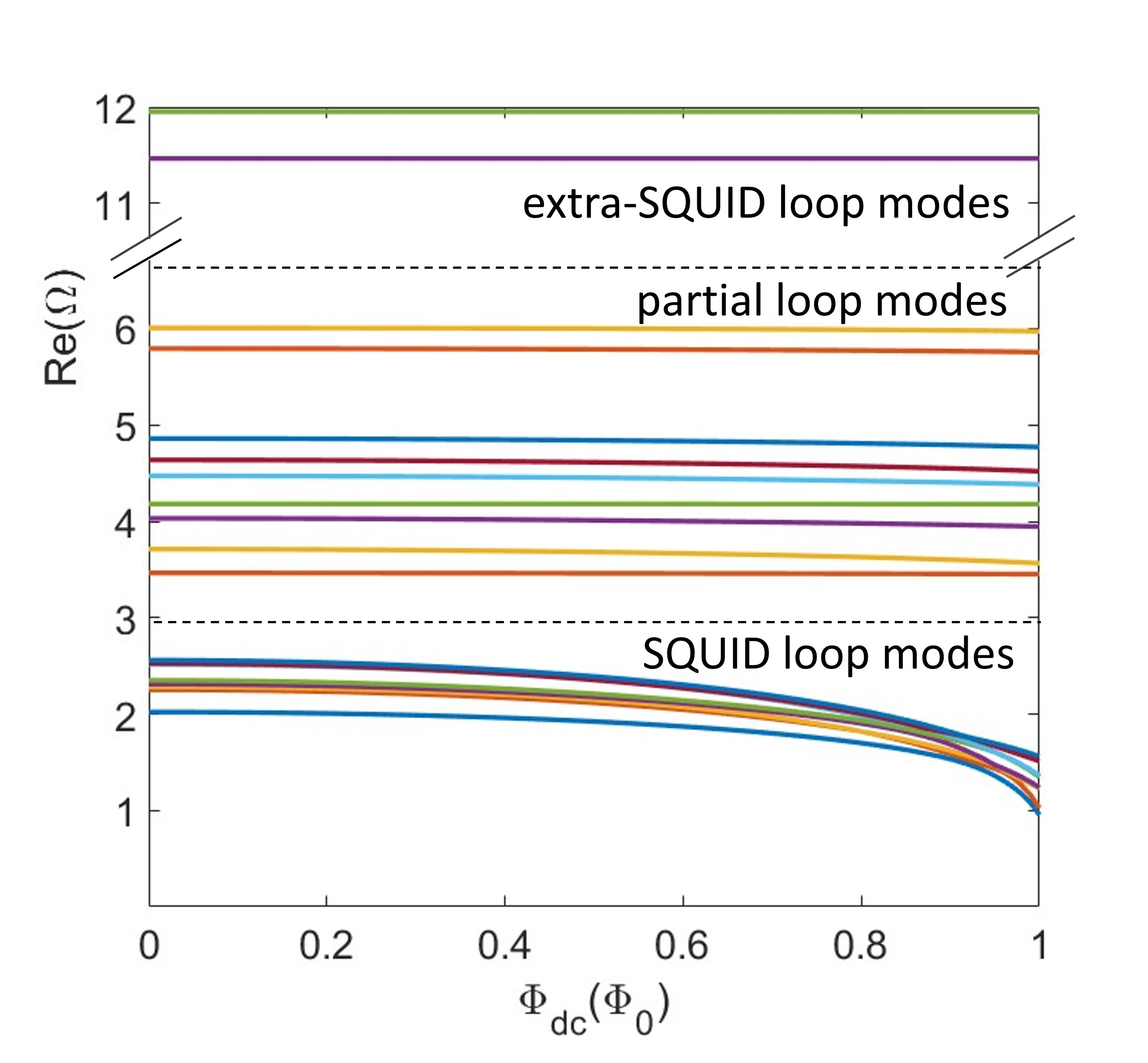}
    \caption{Real part of eigenfrequency solutions $\text{Re}(\Omega)$ from the characteristic equation $\det(\overleftrightarrow\chi)=0$ in the linear limit for the $2\times2\times2$ corner-coupled SQUIDs shown in Fig. \ref{fig:large_SQUID_schem}(b), as a function of dc magnetic flux $\Phi_\text{dc}$ in units of $\Phi_0$. A total of $19$ distinct resonance modes can be resolved. The two black horizontal dashed lines delineate the three types of resonant modes. }
    \label{fig:2x2x2_eig}
\end{figure}

\begin{figure}[H]
    \centering
    \includegraphics[width=0.9\linewidth]{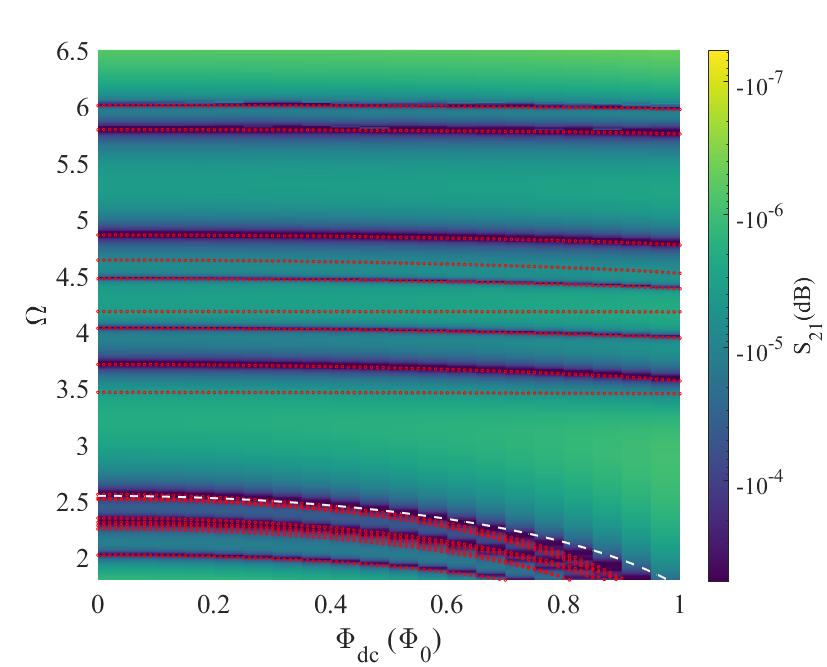}
    \caption{Nonlinear solutions to the $2\times 2\times 2$ system shown in Fig. \ref{fig:large_SQUID_schem}(b) at $\Phi_\text{rf}\sim10^{-3}\Phi_0$. The quantity plotted is $|S_{21}| (\text{dB})$ on a logarithmic color scale, as a function of dimensionless frequency $\Omega= \omega/\omega_\text{geo}$ and applied dc magnetic flux in units of $\Phi_0$. The red dotted curves are the eigenfrequencies from the linear-limit solution matching the resonances from the numerical solution. The dashed white curve is the eigenfrequency of the single SQUID with the same design parameter.}
    \label{fig:2by2by2_nlin}
\end{figure}

Following the same treatment as in Sec. \ref{four_corner_cpl_mdl}, the eigenfrequencies in the low rf flux amplitude linear limit, and the full nonlinear numerical solution, can be obtained, and are shown in Figs. \ref{fig:2x2x2_eig} and \ref{fig:2by2by2_nlin}, respectively. Indeed, $19$ eigenfrequencies fall  in the expected range for their corresponding loop modes as depicted in Fig. \ref{fig:2x2x2_eig}.  We note once again \textcolor{black}{the} general \textcolor{black}{decreasing sensitivity} to dc flux with increasing frequency of the modes, \textcolor{black}{consistent with Eq.(\ref{eqn:resonance_lin_model_Leff})}.

The full numerical solution in the low rf flux amplitude limit in Fig. \ref{fig:2by2by2_nlin} shows resonances coincident with the eigenfrequencies obtained from the linear-limit solution. 

\subsubsection{\texorpdfstring{$N\times N\times 2$}{NbyNby2} corner-coupled overlapping SQUIDs}\label{NbyNby2_mdl}

In the general case of an $N\times N\times 2$ SQUID array, there are $2N^2$ SQUID loops, $(2N-1)^2$ partial loops, and $2(N-1)^2$ extra-SQUID loops, resulting in a total number of $8N^2-8N+3$ distinct loops. As illustrated above, each loop corresponds to an equation of motion in the voltage formalism. Thus, one would expect $8N^2-8N+3$ equations for an $N\times N\times 2$ SQUID array. There are two capacitor nodes for each partial loop. However, the current conservation law from the SQUID loops will constrain the number of independent nodal voltages to $2\times\# \text{ of partial loops}-(\# \text{ of SQUID loops}-1)=6N^2-8N+3$. Together with the $2N^2$ gauge-invariant phase differences $\delta$ for each SQUID, the dynamics is described by a total of $8N^2-8N+3$ variables. The resulting $(8N^2-8N+3)\times(8N^2-8N+3)$ system can be solved exactly (see Appendix \ref{app:general_volt_form_ovlp_SQUIDs}) and will generate $8N^2-8N+3$ modes in the absence of any symmetry.

Equipped with the loop-mode correspondence established in the analysis for the three model systems discussed in section \ref{Theory}, we can extrapolate to the prediction for a larger system, where many partial loop modes should be visible between $\Omega=3 $ and $6$, with very weak dc flux tunability, while the SQUID loop modes occur at lower frequency range with high dc flux tunability. The extra-SQUID loop modes around $\Omega=12$ is almost invariant in dc flux and is beyond the frequency range of our measurement capability.

\section{Experiment}
\label{Expt}
\subsection{SQUID samples}
The samples in this work are fabricated by the STAR Cryoelectronics Selective Niobium Anodization Process (SNAP) with Josephson junction critical current density $J_c=1\mu \text{A}/\mu \text{m}^2$ \cite{Cantor05}. The \ch{Nb} films have a critical temperature of $T_c = 9.2 K$, and the samples were measured at $T = 4.6K$. A representative three-dimensional rf SQUID metamaterial sample is shown in the inset of Fig. \ref{fig:exp_schematics}. The main concern in the design is to maintain the rf-SQUID self-resonant frequency in the 10-30 GHz range while keeping $\beta_\text{rf}=L_\text{geo}/L_\text{JJ}$ reasonably small $(\lesssim10)$ to avoid extreme hysteresis. The low resonant frequency of the rf-SQUID requires a large junction inductance, corresponding to a small critical current.  The requirements of small junction size and low critical current density is an uncommon constraint for Josephson junction foundries, since most superconducting electronics applications prefer high Josephson energy ($E_{J}=\Phi_0I_c/(2\pi)$) and thus high critical current of the junction. 

\subsubsection{SNAP fabrication}
SNAP fabrication starts by first depositing the \ch{Nb}/\ch{Al}-\ch{AlO_x}/\ch{Nb} trilayer on the wafer with $J_c=1\mu \text{A}/\mu \text{m}^2$. This trilayer is then patterned with a wet etch that defines the base layer of the SQUID loop, the vias between the top and the base wiring layers of the SQUID, as well as the anodization rails connecting all of the SQUIDs to the edge of the wafer.  The anodization rails are required to supply a voltage bias to the Nb layers for the anodization process in the next step. The junctions, anodization rails, and the area surrounding the vias \footnote{A via is required  to electrically connect the top and base wiring layers of the SQUID to establish a galvanically-connected loop. Therefore, in SNAP fabrication, one needs to etch through the trilayer to expose the edge of the base \ch{Nb} layer and to protect the etched area with photoresist during the following anodization process. To ensure a good electrical contact and avoid introducing unintended small junctions with low critical current and high junction inductance, a large area ($\approx60\times$ area of the junction) surrounding the via should be protected with photoresist to form a very large-area junction.  This large junction will act as a via, and will not otherwise affect the operation of the rf SQUID, at least at low rf power.} are protected by photoresist, while the remaining exposed Nb is anodized to form the $100\ \text{nm}$-thick insulating dielectric layer of \ch{Nb_2O_5} between the top and base Nb layers.  Previous measurements of the dielectric constant of \ch{Nb_2O_5} made by this process yield $\epsilon_r = 29$ \cite{Basa76,Henkels77}. Next, the second \ch{Nb} layer is deposited and patterned to form the top wiring layer of the SQUID. After the SQUIDs are defined, the anodization rails connecting the SQUIDs are finally removed with a wet etch.

\subsubsection{SNEP fabrication}
The first generation of the overlapping SQUID samples, as shown in Fig. \ref{fig:corner_schem}, was made with a slightly more complicated process, the Selective Niobium Etch Process (SNEP) from STAR Cryoelectronics. This process begins with the same \ch{Nb}/\ch{Al}-\ch{AlO_x}/\ch{Nb} trilayer deposition. However, instead of anodization, the junctions are defined by a reactive ion etch (RIE) on the top \ch{Nb} layer of the trilayer. The entire trilayer is then patterned and etched to form the base wiring layer. A layer of \ch{SiO_2} with a thickness of $300\ \text{nm}$ is then deposited with plasma enhanced chemical vapor deposition (PECVD), and this serves as the insulating dielectric between the top and base layers. The dielectric deposition is then followed by another etch to open contact vias through the \ch{SiO_2} layer. Next, the top \ch{Nb} wiring layer is deposited and patterned. The process concludes with a final passivation of the wafer by depositing a layer of \ch{SiO_2}. In addition to its complexity, the unit area capacitance from the dielectric \ch{SiO_2} in SNEP is very low $\approx 0.15\ \text{fF}/\mu \text{m}^2$, compared to that for the \ch{Nb_2O_5} dielectric in SNAP $\approx 2.5\ \text{fF}/\mu \text{m}^2$. Therefore, to maintain a low self-resonance for the rf SQUID, a larger capacitor pad is needed for samples made by the SNEP process, which further constrains the design.

\subsubsection{SQUID array design}
Since the fabrication processes only allow for a single trilayer for the junction definition, one cannot simply stack two independently-defined layers of 2D rf SQUID arrays in the third dimension.  One of the layers must be shifted in-plane so that its junction avoids that of the other layer of SQUIDs, and this constraint creates the peculiar overlapping geometry studied here. To achieve the most symmetric configuration, the overlapping area between the SQUIDs from the two layers is designed to be roughly a quarter of the single-SQUID loop area.  The shifted stack between the two 2D rf SQUID arrays results in the many overlapping capacitors $C_\text{ov}$ between SQUID loops from different layers. The design parameters for the overlapping SQUID metamaterial sample SNAP 161A is summarized in Table.\ref{tab:SQUIDparams}

\begin{table}[H]
    \centering
    \begin{tabular}{|c|l|c |}\hline
    Parameters& Symbols&Values\\\hline  
Gap between nearest SQUID wiring&   $d$&$4\ \mu \text{m}$\\ \hline 
 Wiring width&   $w$&$16\ \mu \text{m}$\\ \hline 
SQUID loop side length&   $a_\text{SQUID}$&$132\ \mu \text{m}$\\ \hline 
 SQUID self capacitance (RCSJ)&   $C$&$1.42\ \text{pF}$\\ \hline 
 Overlapping capacitance& $C_\text{ov}$&$0.657\ \text{pF}$\\ \hline
  Junction critical current ($4.2 K$)&   $I_c$&$7\ \mu \text{A}$\\ \hline 
    Geometric Inductance&   $L_\text{geo}$&$255.2\ \text{pH}$\\ \hline
     SQUID parameter $L_\text{geo}/L_\text{JJ}$&   $\beta_\text{rf}$&$5.483$\\ \hline
 \makecell{Mutual inductance between \\ the overlapping neighbors}& $M_1$ &$8.56\ \text{pH}$\\\hline
 \makecell{Mutual inductance between \\ the in-plane nearest neighbors}& $M_2$&$-17.4\ \text{pH}$\\\hline
    \end{tabular}
    \caption{Design parameters for the overlapping SQUID metamaterial sample SNAP 161A.  See Fig. \ref{fig:corner_schem}(b) for definitions of $d$, $w$, and $a_\text{SQUID}$.} 
    \label{tab:SQUIDparams}
\end{table}

The mutual inductances in the last two rows of the Table \ref{tab:SQUIDparams} represent the inductive coupling in the absence of the overlapping capacitors. However, under the strong capacitive coupling studied in this work, the rf currents can leave the SQUID loop through the capacitor nodes, breaking the uniformity of the current within one SQUID loop. The mutual inductance alone is thus no longer sufficient to correctly treat the coupling between SQUIDs.

\subsection{Experimental Setup}
\begin{figure}[H]
    \includegraphics[width=1\linewidth]{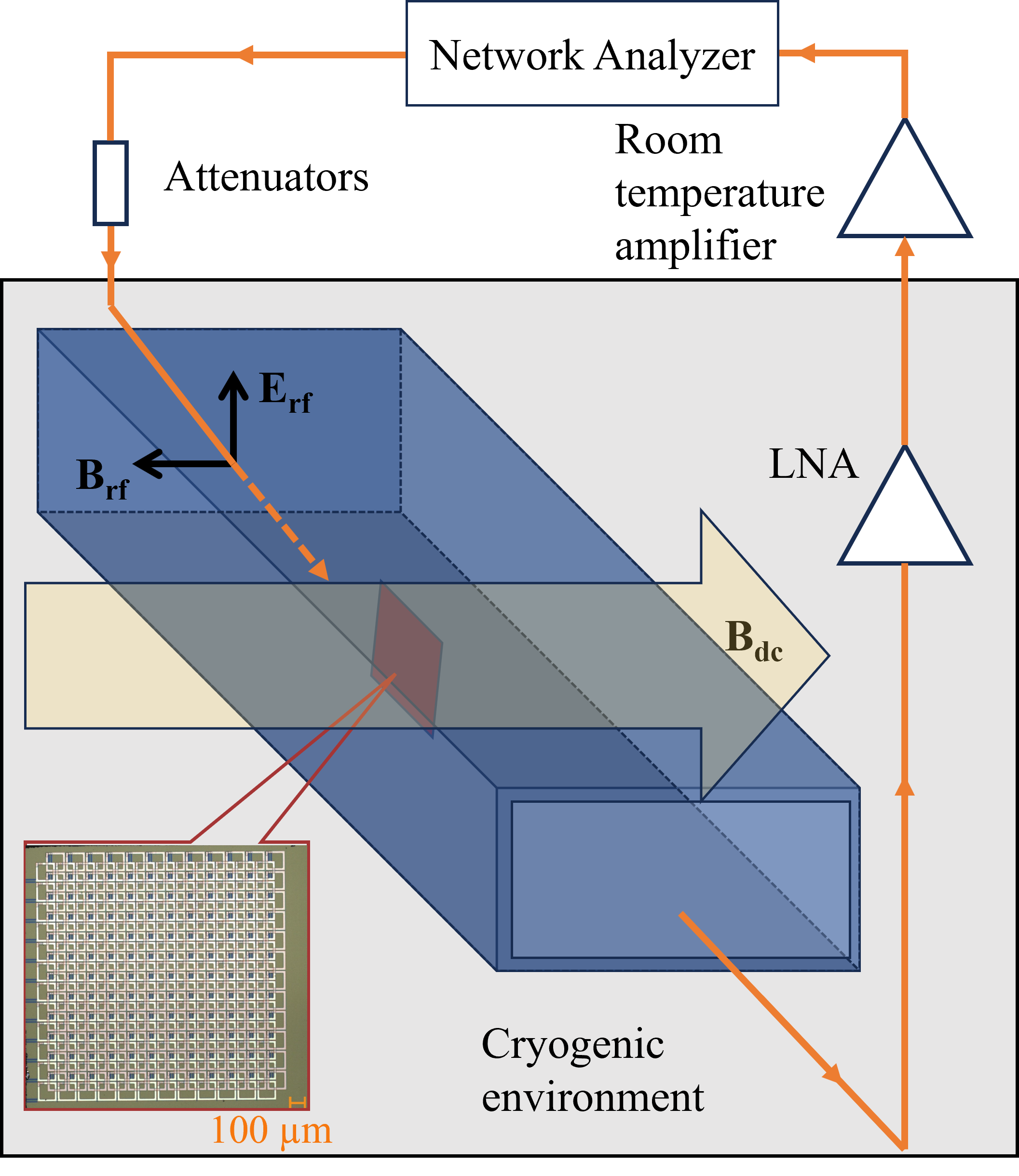}
    \caption{Schematic of the rf SQUID metamaterial transmission measurement, showing the waveguide which hosts the metamaterial, the rf and dc field components, the microwave network analyzer, as well as attenuators and amplifiers. Inset:  optical image of the overlapping 12$\times$12$\times$2 SQUID array sample made with the SNAP process. The blue shaded links between SQUIDs are the remnants of the anodization rails after wet etch. Variables that can be controlled in the experiment include $\Phi_\text{rf}$, $\Phi_\text{dc}$, rf driving frequency $f$, and sample temperature $T$.}
    \label{fig:exp_schematics}
\end{figure}

The rf SQUID metamaterial is embedded in the center of a brass WR-42 waveguide, which has a cut-off frequency of approximately 14.1 GHz, and provides good transmission above 15 GHz. The sample is oriented along the direction of the wave propagation and its planar surface is perpendicular to the rf magnetic field in the first propagating TE mode of the waveguide. The dc magnetic flux is provided by a superconducting Helmholtz coil mounted on the outside of the waveguide and oriented to apply a nominally uniform field perpendicular to the SQUID array surface. The microwave transmission $S_{21}$ through the sample-loaded waveguide is measured by a  microwave vector network analyzer (VNA), as shown in Fig. \ref{fig:exp_schematics}.  The amplitude of the incident signal is reduced by input attenuators to carefully control the rf flux $\phi_\text{rf}$ witnessed by the sample.  The transmitted signal is amplified by a cryogenic amplifier and by a room temperature amplifier, before being measured by the VNA. 
 The measurements are carried out at a temperature of 4.6 $K$.  Note that no wires are connected to the rf SQUIDs, and there are no galvanic contacts between different SQUIDs,  hence all changes to the properties of the metamaterial occur by strictly `wireless' means.  

\subsection{Measurement results}  

\begin{figure}[H]
    \includegraphics[width=\columnwidth]{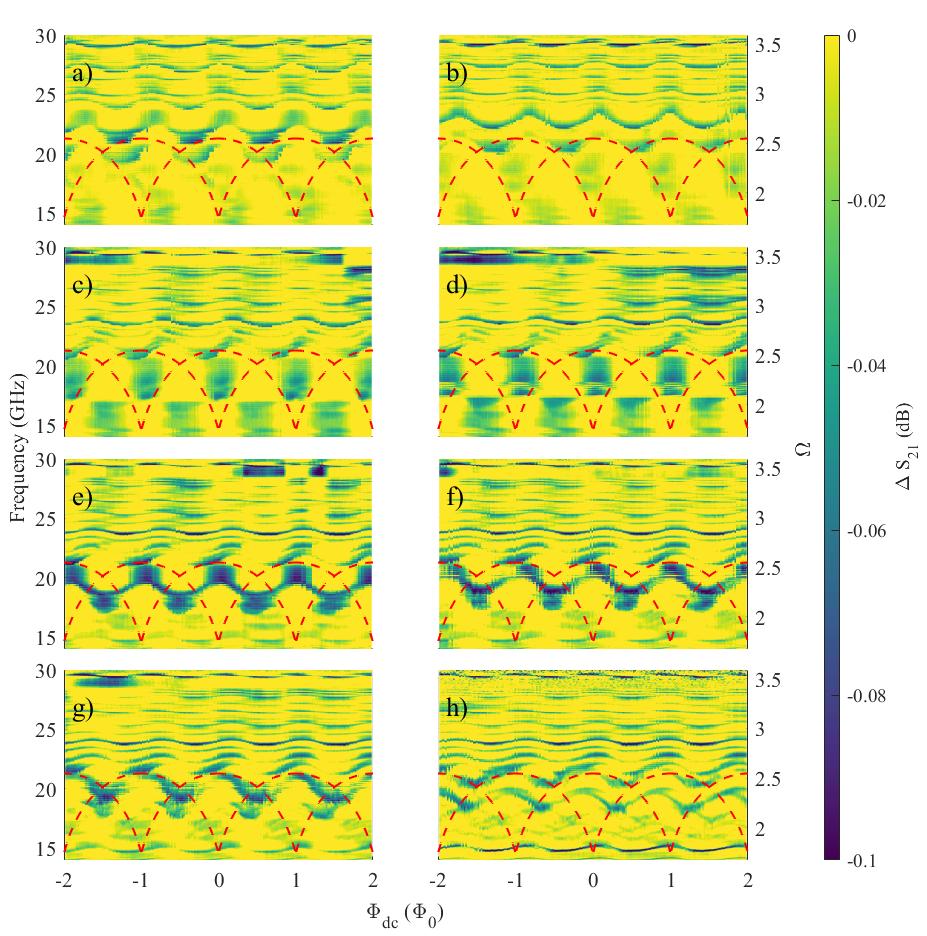}
    \caption{Measured change in transmission $\Delta S_{21}$ (blue to yellow color) of the $12\times12\times2$ overlapping corner-coupled SQUID array as a function of dc flux swept from $-2$ to $+2\ \Phi_0$, and as a function of rf driving frequency $f$ from 15 to 30 GHz ($\Omega = f/f_\text{geo} = 1.8 - 3.6$), at a temperature of 4.6 $K$.  Results are presented at eight different applied rf flux amplitudes: $[0.033,\ 0.023,\ 0.013,\ 0.01,\ 0.007,\ 0.006,\ 0.005,\ 0.0006]\ \Phi_0$ in panels a) through h). The red dashed curves denote the single-SQUID eigenfrequency.  }
    \label{fig:exp_2layer_array_rfdep}
\end{figure}

The rf SQUID response is typically very small in the transmission measurement due to its weak coupling to the waveguide mode. To better resolve the SQUID resonances, a background response obtained from averaging over the transmission spectra $\overline{S}_{21}$ at different dc fluxes is removed from the individual frequency spectrum. The resulting $\Delta S_{21} = S_{21}(f, \Phi_\text{dc}) - \overline{S}_{21}$ as a  function of applied dc flux $\Phi_\text{dc}$ and microwave frequency $f$ is plotted in Fig. \ref{fig:exp_2layer_array_rfdep} for eight different values of the rf flux amplitude $\Phi_\text{rf}$.  

The yellow color represents nearly complete transmission of the microwaves (i.e. $\Delta S_{21} \lesssim 0 \text{ dB}$), while the darker green features show conditions where the metamaterial interacts strongly with the passing electromagnetic fields and dissipates power.  The green features trace out the resonant response as a function of applied dc magnetic flux with a typical $1\Phi_0$ periodicity. The red dashed curves mark the single-SQUID resonance and roughly corresponds to the expected  boundary between the low-frequency and highly-tunable SQUID modes, and the high-frequency partial loop modes with lower-tunability introduced in Sec. \ref{four_corner_cpl_mdl}. It should be noted that the third kind of mode, the extra-SQUID loops modes, with frequencies around $12 f_\text{geo}\approx 100 \text{ GHz}$, are beyond the measurement range of our apparatus. 

\begin{figure}[H]
    \centering
    \includegraphics[width=1\linewidth]{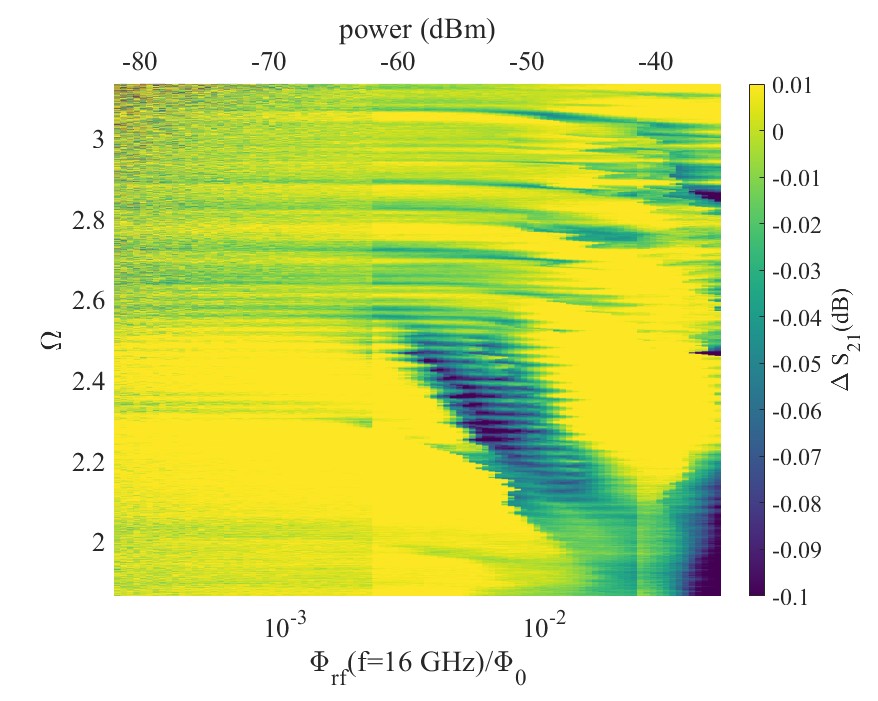}
    \caption{Measured change in transmission $\Delta S_{21}$ (green to yellow color) of the $12\times12\times2$ overlapping corner-coupled SQUID array as a function of applied rf magnetic flux amplitude from $2\times 10^{-4}\Phi_0$ to $5\times 10^{-2} \Phi_0$ (and the corresponding microwave power incident on the sample in dBm), and rf driving frequency $f$ from 15 to 26.5 GHz ($\Omega =1.8\ \text{to}\ 3.2$), at a temperature of 4.6 $K$. The measurement was taken at $\Phi_{dc}=-0.3 \Phi_0$ with two separate power sweeps due to the limited 40 dB dynamic range in the VNA power sweep function. The two vertical lines around -62 and -42 dBm are the artifacts from stitching the two sweeps together.}
    \label{fig:exp_rf_dep_snap161A}
\end{figure}

Figure \ref{fig:exp_rf_dep_snap161A} shows the effects of increasing the driving rf flux amplitude beyond linear response on the spectrum of modes in the overlapping $12\times12\times 2$ sample.  The measurement was performed at zero current in the magnet but $\Phi_\text{dc}=-0.3 \Phi_0$ on the SQUIDs, based on their dc flux tunability curves. The result in Fig. \ref{fig:exp_rf_dep_snap161A} is obtained from two separate power sweeps on the VNA due to its limited dynamic range. The first power sweep was performed from -82 to -42 dBm, while the second from -62 to -35 dBm. In the range where both power sweeps overlap, the average response is shown in Fig. \ref{fig:exp_rf_dep_snap161A} leading to the two vertical lines at the limits of the two power sweeps. 

One notes strong tuning of the modes below $\Omega = 2.5$ and relatively small tuning for the higher frequency modes.  This behavior is in qualitative agreement with that shown in Fig. \ref{fig:nonline_sol_2_corner_rf} for the two corner-coupled SQUIDs.  The low-frequency SQUID loop modes are strongly tuned, whereas the partial-loop modes only show modest tuning with rf flux.  The experiment appears to just reach the high-power linear limit that is clearly seen in the model results in Fig. \ref{fig:nonline_sol_2_corner_rf}.  However, these large rf flux amplitudes bring about the dangers of sample heating and amplifier saturation.

\begin{figure}[H]
    \centering
    \includegraphics[width=1\linewidth]{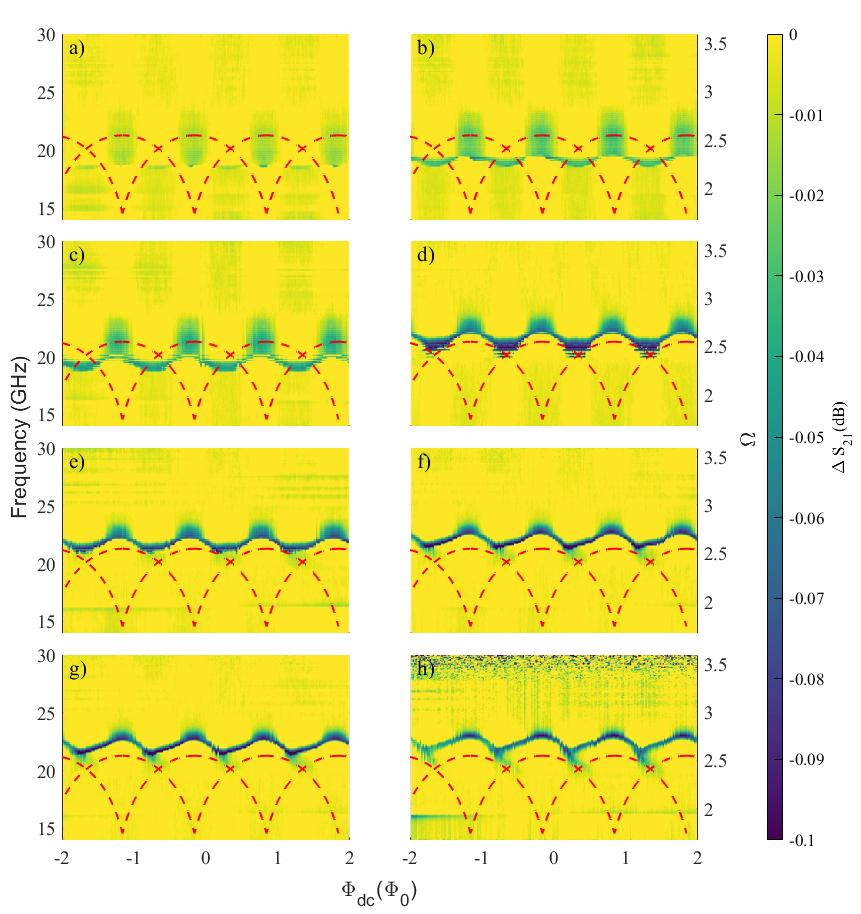}
    \caption{Measured change in transmission $\Delta S_{21}$ (blue to yellow color) of the $12\times12\times 1$ single-layer SQUID array as a function of dc flux swept from $-2$ to $+2\ \Phi_0$, and as a function of rf driving frequency $f$ from 15 to 30 GHz ($\Omega = 1.8-3.6$), at a temperature of 4.6 $K$.  Results are presented at eight different applied rf flux amplitudes: $[0.042,\ 0.03,\ 0.026,\ 0.017,\ 0.015,\ 0.011,\ 0.009,\ 0.0003]\ \Phi_0$ in panels a) through h). The red dashed curves denote the single-SQUID eigenfrequency.}
    \label{fig:exp_1layer_array_rfdep}
\end{figure}

The response from the overlapping SQUID array SNAP161A is in clear contrast to the measurement on a single layer $12\times12\times 1$ SQUID array (SNAP161D) shown in Fig. \ref{fig:exp_1layer_array_rfdep}. \textcolor{black}{The single-layer sample shares the same design with the bottom layer of the two-layer sample (SNAP161A).} Unlike the overlapping corner-coupled SQUID array, the single layer SQUID array has only one resonance band tuned with the applied dc flux $\Phi_\text{dc}$, which is consistent with the coherent response seen in earlier measurements of single-layer SQUID metamaterials \cite{Trep13,Trep17,Zhu19,Zack22}.  The resonant frequency is close to that of a single-SQUID resonance shown as the red dashed curves.  At low applied rf flux amplitude, the resonance is above that of the single SQUID since the collective antiferromagnetic coupling reduces the induced flux on each SQUID and thus the effective inductance. According to Eq.(\ref{eqn:resonance_lin_model_Leff}), a low $L_\text{eff}<L_\text{geo}$ corresponds to a resonance higher than that of the single SQUID as observed experimentally in Fig. \ref{fig:exp_1layer_array_rfdep}. As expected for the SQUID modes, their tunability in $\Phi_\text{dc}$ also diminishes with increasing $\Phi_\text{rf}$, again observed in earlier work on single-layer SQUID metamaterials \cite{Trep13,Trep17,Zhu19,Zack22}.  

There are several other interesting features of Fig.  \ref{fig:exp_1layer_array_rfdep} worth mentioning.  First, note that green blobs appear around the turning points where the resonance tuning curves in applied dc flux from the two adjacent periods meet.  These can be attributed to the fact that the meta-atom SQUID is strongly hysteretic ($\beta_\text{rf}>1$), resulting in multiple stable solutions.  At points where multiple solutions cross in the frequency-dc-flux space, there can be enhanced resonant responses over a range of frequencies, accounting for these blobs.  We also note that the tuning curves are noticeably asymmetric as a function of dc flux at low rf flux amplitudes, and become increasingly symmetric as the rf flux amplitude increases. The asymmetry in the dc tuning curve is related to the hysteresis in dc flux sweep of the rf SQUID metamaterial. For a measurement with decreasing dc flux, the asymmetric tilt of the tuning curve points to the opposite direction. This hysteresis is suppressed at higher applied rf flux amplitude because the strong oscillatory drive can overcome the barriers between local potential minima, preventing the system from becoming stuck in metastable states that are responsible for the hysteresis.

\section{Discussion}
\label{Disco}
Two driven side-by-side inductively-coupled rf SQUIDs have two eigenmodes of oscillation in the linearized case.  When the two loops are partially overlapping with significant capacitive coupling, a third eigenmode of oscillation appears, which arises from the partial loop created by the partial overlap between the SQUID loop wires. This new closed circuit is completed by means of displacement currents. The same loop-to-oscillation-mode correspondence has been verified in the calculations for larger systems, i.e. four corner-coupled SQUIDs and the two by two by two array of overlapping SQUIDs.

\begin{figure}[H]
    \centering
    \includegraphics[width=1\linewidth]{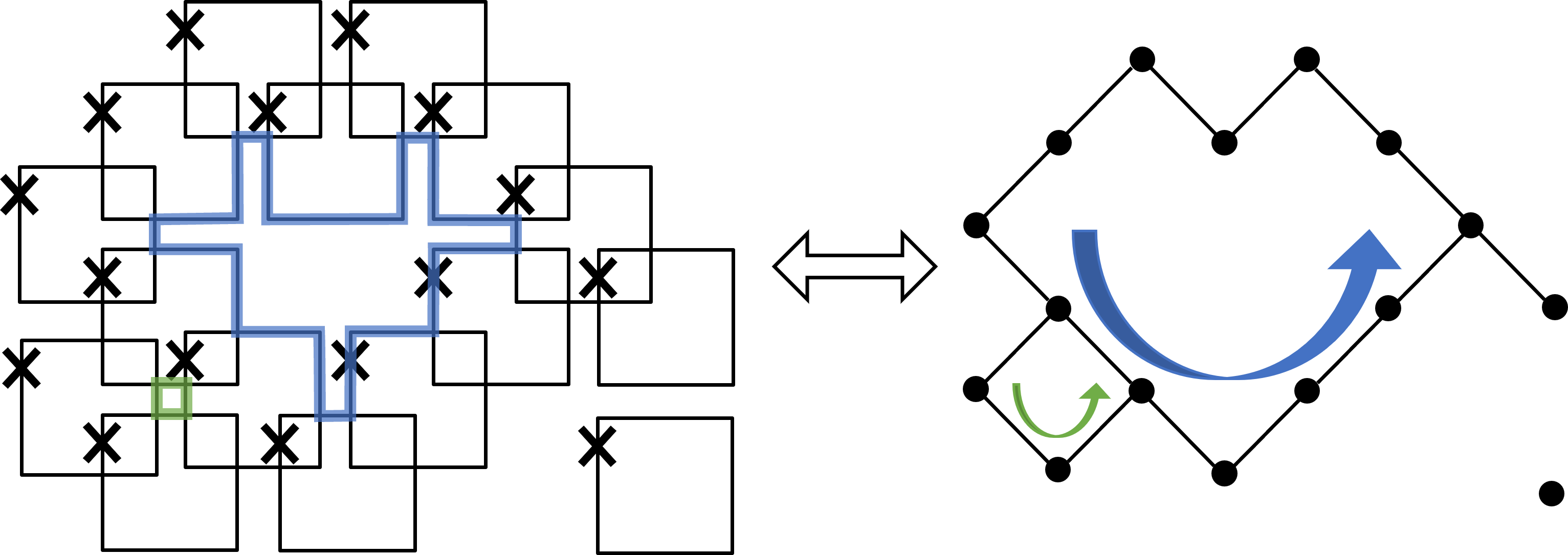}
    \caption{The correspondence between a general system of overlapping SQUIDs and its graph representation. The SQUID loops are represented by the vertices, and the partial loops connecting the neighboring SQUIDs by the edges. The simple cycles in the graph correspond to extra-SQUID loops in the SQUID array, as highlighted by green and blue arrows in the graph, and corresponding closed loops in the SQUID array.  }
    \label{fig:graph_SQUID_illustration}
\end{figure}

More generally, for a system of many overlapping \textcolor{black}{corner-coupled} SQUIDs, the total number of elementary loops can be determined as follows. If we treat the SQUID loops as vertices and partial loops as edges connecting the corresponding vertices, we can represent the geometry of overlapping SQUIDs as a planar graph, as illustrated in Fig. \ref{fig:graph_SQUID_illustration}. 

The most general system can be represented \textcolor{black}{as} a disconnected graph, as shown in Fig.\ref{fig:graph_SQUID_illustration} where a single SQUID in the lower right corner does not overlap with any other SQUID. Without loss of generality, we can examine each connected sub-graph individually and sum up the number of elementary loops. For each connected planar sub-graph with $n$ vertices (SQUID loops) and $m$ edges (partial loops), \textcolor{black}{the simple cycles of the graph correspond to the extra-SQUID loops}, the number of which is given by $m-n+1$, bringing the total number of elementary loops to $m+n+m-n+1=2m+1$ in the sub-graph. The number of unknowns in the voltage formalism for the sub-graph can be obtained as in Sec. \ref{NbyNby2_mdl}, \textcolor{black}{where $2m-n+1$ unknown voltages and} the $n$ phase differences from $n$ junctions, \textcolor{black}{give rise to} $2m+1$ unknowns, equal to the number of elementary loops. \textcolor{black}{Tallying up} all the sub-graphs, we have shown that for any general system of overlapping SQUIDs, the voltage formalism can result in an exactly determined system that describes the dynamics of the SQUIDs. For example, a single SQUID by itself is a planar connected graph with 1 vertex and 0 edges. The dynamics are described by 1 unknown, the phase difference, and 1 equation of motion, the flux quantization condition for a single SQUID (Eq. (\ref{eqn:FluxEq})). The $N\times N\times2$  system in Sec. \ref{NbyNby2_mdl} is another special case of a system consisting of one planar connected graph.

Although, the extra-SQUID loops studied in the examples above are the smallest circuits corresponding to the very high frequency modes $12 f_\text{geo}\approx 100 \text{ GHz}$, they can take on larger sizes in a more general geometry \textcolor{black}{as shown in Fig.\ref{fig:graph_SQUID_illustration}.} In fact, the extra-SQUID loop highlighted in blue in Fig.\ref{fig:large_SQUID_schem} is the smallest realization permitted in our overlapping SQUID design.

The frequency and the tunability in applied dc magnetic flux of the resonance modes can be obtained from Eq.(\ref{eqn:resonance_lin_model_Leff}), where the effective inductance depends on the current distribution in the corresponding resonance modes. The modes with very small $|L_\text{eff}| \ll L_\text{geo}$ will have a high resonant frequency $\Omega_\text{res} \approx \sqrt{L_\text{geo}/L_{a,\text{eff}}}$, and be largely independent of dc flux applied to the metamaterial.  On the other hand, the modes with large effective inductance $|L_\text{eff}|\gg L_\text{geo}$ will have a low resonant frequency $\Omega_\text{res} \approx \sqrt{\beta_\text{rf} \cos \delta}$, even lower than the single SQUID resonance, and will be strongly tunable with dc flux.  The experimental data on dc-flux tuning of the $12\times 12\times 2$ metamaterial is consistent with these expectations.

It should be noted that the only source of loss treated in this work arises from quasiparticle tunneling, which appears as the resistance $R$ in the RCSJ model, and is incorporated into the fully nonlinear equations of motion through the parameter $\gamma$.  An extension of this treatment would be inclusion of other sources of loss.  One candidate source of loss arises from dielectric in the Josephson junction tunnel barrier, the coupling capacitors between SQUIDs, as well as dielectrics surrounding the superconducting wiring.  These dielectrics are known hosts for electric-dipole two-level systems \cite{and72,Phil72}.

The treatment presented here assumes that all of the rf SQUIDs making up the metamaterial are nominally identical, having the same values of geometrical inductance, shunt capacitance, overlap capacitance, critical current, and junction resistance.  We also assume that every SQUID experiences the same values of the externally-applied dc and rf flux, and that the driving rf flux is at a single frequency.  It would be interesting to see how the results of this work depend on variations in these quantities due to either statistical or systematic variation in space.

\textcolor{black}{Although our model has mainly focused on the low rf flux amplitude linear limit, we can explore the nonlinear properties of the system at higher rf flux amplitudes from the full numerical solutions to the equations of motion. Thanks to the plethora of resonance bands in the overlapping SQUID metamaterial, }one can employ them for broadband parametric amplification or intermodulation generation \textcolor{black}{by harnessing this nonlinearity from the Josephson junctions}.

Now that capacitive coupling between flux-based superconducting meta-atoms has been established, one can ask whether the capacitive coupling can be varied?  For example, there exist many nonlinear dielectric materials whose dielectric properties can be tuned with dc electric field, rf electric field, or with temperature at cryogenic \textcolor{black}{conditions} \cite{Tunable_Dielectrics_2010,Alp94}. In addition, charge qubits can enjoy variable capacitive coupling through a small Josephson junction \cite{Averin03}.  Thus a certain degree of tunability of capacitive coupling should be possible.

\section{Conclusions}

We consider three-dimensional rf SQUID metamaterials with strong capacitive coupling between 
rf SQUID loops for the first time. Strong displacement currents can flow through the capacitors 
created by such overlap of the SQUID loops, creating new closed paths for the rf current through the rf SQUID network. The RCSJ model is extended to incorporate the capacitive coupling and the new current paths, leading to the prediction of multiple resonances of comparable frequency to those of the galvanically-closed SQUID loops.  The number of resonating loops in our three-dimensional $N\times N\times 2$ rf SQUID metamaterial design scales as $8N^2 -8N+3$.  A large $N=12$ three-dimensional rf SQUID metamaterial is measured, and is found to behave in a qualitatively different manner from the corresponding single-layer $12\times 12\times 1$ metamaterial. The observed multiplicity of resonances are in good agreement with our theory of capacitively-coupled overlapping SQUIDs.

\section{Acknowledgements}
This work was supported by ONR under grant N000142312507, and NSF/DMR under grant 2004386.

\section{Data availability statement}
The data that support the findings of this study are openly available with the following DOI: https://doi.org/10.13016/dspace/rit2-egnl.

\clearpage

\appendix
\section{Parameters for the two corner-coupled SQUIDs}
\label{app:param_def}
To simplify the discussion in the main text, the definitions and numerical values of the parameters used in Sec.\ref{two_corner_cpl_mdl} are summarized below.

\begin{eqnarray*}
    \kappa_1=M/L_\text{geo}=0.0335\\
    \overleftrightarrow\kappa=\begin{pmatrix}
        1&\kappa_1\\\kappa_1&1
    \end{pmatrix}\\
    L_{\delta a}=L_{a,a1}-M_{a,b1}=39.21 \text{pH}\\
    L_{\delta b}=L_{b,b1}-M_{b,a1}=207.39 \text{pH}\\
    \kappa_{\delta a}= L_{\delta a}/L_\text{geo}=0.154\\
    \kappa_{\delta b}= L_{\delta b}/L_\text{geo}=0.813\\
    \kappa_a = \frac{L_\text{geo}}{L_{\delta a}}+\frac{M}{L_{\delta b}}=6.55\\
    \kappa_b = \frac{M}{L_{\delta a}}+\frac{L_\text{geo}}{L_{\delta b}}=1.45\\    
    \text{CD} = \det{
    \begin{pmatrix}
        L_{a,a1}& M_{a,b1}\\
        M_{b,a1}& L_{b,b1} 
    \end{pmatrix}}\\
    \kappa_{\text{v}a} = \det{\begin{pmatrix}
        M_{\text{cen},b1}& L_{\text{cen},a1}\\
        L_{b,b1}& M_{b,a1} 
    \end{pmatrix}}/\text{CD}=-0.714\\
     \kappa_{\text{v}b} = \det{\begin{pmatrix}
        M_{a,b1}& L_{a,a1}\\
        M_{\text{cen},b1}& L_{\text{cen},a1} 
    \end{pmatrix}}/\text{CD}=-0.0085\\
    L_{\text{I}a}=\det{
    \begin{pmatrix}
        M_{\text{cen},a0}& L_{\text{cen},a1}& M_{\text{cen},b1}\\
        L_{a,a0} & L_{a,a1}& M_{a,b1}\\
        M_{b,a0}& M_{b,a1}& L_{b,b1}
    \end{pmatrix}}/\text{CD}=-134.4 \text{pH}\\
    L_{\text{I}b}=\det{
    \begin{pmatrix}
        L_{\text{cen},b0}& L_{\text{cen},a1}& M_{\text{cen},b1}\\
        M_{a,b0} & L_{a,a1}& M_{a,b1}\\
        L_{b,b0}& M_{b,a1}& L_{b,b1}
    \end{pmatrix}}/\text{CD}=39.5 \text{pH}\\
    \kappa_{\text{I}a}=L_{\text{I}a}/L_\text{geo}=-0.527\\
    \kappa_{\text{I}b}=L_{\text{I}b}/L_\text{geo}=0.155\\
    \overleftrightarrow{\kappa}_\text{I}=\begin{pmatrix}
        \kappa_{\text{I}a}& \kappa_{\text{I}b}\\
        \kappa_{\text{I}a}& \kappa_{\text{I}b}
    \end{pmatrix}\\
    \lambda_\text{cov}=C_\text{ov}/(2C)=0.23\\
    \overleftrightarrow{\kappa}_\delta=
    \begin{pmatrix}
        \kappa_{\text{v}a}& 1+\kappa_{\text{v}b}\\
        \kappa_{\text{v}a}& 1+\kappa_{\text{v}b}
    \end{pmatrix}\\
    \overleftrightarrow{\kappa}_\text{loop}=
    \begin{pmatrix}
        -\kappa_{\delta a}& 0\\
        0& \kappa_{\delta b}
    \end{pmatrix}\\
    \alpha=\kappa_b\Phi_\text{cen}^\text{app}/\Phi^\text{app}+\kappa_b\kappa_{\text{v}a}-\frac{L_{\text{I}b}}{ L_{\delta a}}+\kappa_b\kappa_{\text{v}b}-\frac{L_{\text{I}b}}{ L_{\delta b}}=-2.00\\
\end{eqnarray*}
The partial inductance matrix for the 2 corner-coupled SQUIDs (as shown in Fig. \ref{fig:corner_schem}(a)) has the following numerical values 
\begin{eqnarray}
    \begin{pmatrix}
    L_{a,a0}& M_{a,b0}& L_{a,a1}& M_{a,b1}\\
    M_{b,a0}& L_{b,b0}& M_{b,a1}& L_{b,b1}\\
    M_{\text{cen},a0}&L_{\text{cen},a1}&L_{\text{cen},b0}&M_{\text{cen},b1}
\end{pmatrix}=\nonumber\\
\begin{pmatrix}
  203.27& -4.12&51.89& 12.68\\
    12.68&  51.89&-4.12& 203.27\\
    10.77&37&37&10.77
\end{pmatrix} \text{pH}
\end{eqnarray}

\section{Numerical values for the parameters in our SNAP SQUID design}
\label{app:design_parameters}
Here we list the parameters of the basic rf SQUID unit that makes up all of our studies, both numerical and experimental.  The geometric inductance $L_\text{geo}=255.16\ \text{pH}$, and mutual inductance $M=+8.56\ \text{pH}$  are calculated analytically and verified by Fast Henry numerical solution. The positive mutual inductance is due to the partial overlap of the loops in this geometry. The capacitance for the junction pads and the overlapping capacitors are $C=1.42\ \text{pF}$, $C_\text{ov}=0.657\ \text{pF}$, respectively, determined from the geometrical area of the capacitors and the known thickness and composition of the \ch{Nb_2O_5} dielectric material. All the calculations use the parameter $\beta_\text{rf}=L_\text{geo}/L_\text{JJ}=5.483$, determined from the junction critical current $I_c = 7\ \mu A$ measured at 4.2K, while the experiments on the SQUID metamaterials were performed at 4.6K. We also estimated the sub-gap resistance as  $R = 500\ \Omega$, corresponding to $\gamma=0.0268$.  The geometrical resonance frequency for a single SQUID is $f_\text{geo}=\frac{1}{2\pi \sqrt{L_\text{geo} C}}=8.36$ GHz.  Other numerical values for derived quantities in the 2 corner-coupled SQUID case are given in Appendix \ref{app:param_def} and Table \ref{tab:SQUIDparams}.

\section{Voltage formalism for two corner-coupled SQUIDs}\label{app:volt_form_2_corner_cpl}
Due to its simple geometry, the model for two corner-coupled SQUIDs can be completely expressed through the currents in each wiring segment, as discussed in Sec. \ref{two_corner_cpl_mdl}. Here we present the alternative method to set up and solve the equations of motion using the more general voltage formalism.  This approach is convenient for modelling larger arrays of capacitively coupled rf SQUIDs. 

 The junction currents $I_{a,b}$ can be expressed in terms of $\delta_{a,b}$ and $\dot V_1$ from the equation of motion, Eq. (\ref{eqn:eom_corner2}): 

\begin{eqnarray}
\begin{pmatrix}
    I_{a0}\\I_{b0}
\end{pmatrix}
=\overleftrightarrow L^{-1}\nonumber\\
\begin{pmatrix}
C_\text{ov}L_{\delta a}\dot{V}_1-\Phi_0\delta_a/(2\pi)+\Phi_\text{app}\\
-C_\text{ov}L_{\delta b}\dot{V}_1-\Phi_0\delta_b/(2\pi)+\Phi_\text{app}
\end{pmatrix}
    \label{eqn:2crnr-cpl_I_in_volt}\\
    \begin{pmatrix}
    \iota_{a0}\\\iota_{b0}
\end{pmatrix}
=\overleftrightarrow \kappa^{-1}\nonumber\\
\begin{pmatrix}
2\lambda_\text{cov}\kappa_{\delta a}\dot{u}_1-\delta_a+\phi_\text{app}\\
-2\lambda_\text{cov}\kappa_{\delta b}\dot{u}_1-\delta_b+\phi_\text{app}
\end{pmatrix}\nonumber
\end{eqnarray}

where the second equation is the dimensionless form, with $u_1 = 2\pi V_1/(\Phi_0 \omega_\text{geo})$, and $\overleftrightarrow L=L_\text{geo}\overleftrightarrow\kappa=\begin{pmatrix}
L_\text{geo}&M\\M&L_\text{geo}
\end{pmatrix}$. We can then substitute the time derivative of Eq. (\ref{eqn:2crnr-cpl_I_in_volt}) into Faraday's laws, Eq. (\ref{eqn:faraday_center_corner2}) and obtain the expression for $\ddot{u}_1(\dot \delta_a, \dot \delta_b, u_1)$:
\begin{eqnarray*}
   \ddot{u}_1 = [\dot{\phi}_\text{cen}^\text{app}-\dot{\phi}^\text{app}(M_{\text{cen},a}+M_{\text{cen},b})/(L+M)\\\nonumber
   +\dot{\delta}_a(L M_{\text{cen},a}-MM_{\text{cen},b})/(L^2-M^2)+\\\nonumber
   \dot{\delta}_b((L M_{\text{cen},b}-MM_{\text{cen},a})/(L^2-M^2)-1)+2u_1]\\\nonumber
   (2\lambda_\text{cov}/L[(M_{\text{cen},b1}-
L_{\text{cen},a1})+\\\nonumber
((LM_{\text{cen},b}-MM_{\text{cen},a})(M_{b,a1}-L_{b,b1})\\\nonumber
+(LM_{\text{cen},a}-MM_{\text{cen},b})(L_{a,a1}-M_{a,b1}))/(L^2-M^2)])^{-1}
\end{eqnarray*}
where $M_{\text{cen},a}=L_{\text{cen},a1}+M_{\text{cen},a0}$, $M_{\text{cen},b}=L_{\text{cen},b0}+M_{\text{cen},b1}$, and $L_\text{geo}$ is abbreviated as $L$.
Consequently, the equation of motion can be fully expressed in terms of a new set of six variables,  $\delta_{a,b},\ \dot{\delta}_{a,b},\ u_1,\ \text{and }\dot u_1$. One can therefore set up the six first-order initial value problems for the numerical solver in the following manner:
\\
\begin{subequations}
\label{eqn:alt_eom_num_2_corner}
\begin{align}
&\frac{d\delta_a}{d\tau}=\dot{\delta}_a\\
&\frac{d\delta_b}{d\tau}=\dot{\delta}_b\\
&\frac{d\dot{\delta}_a}{d\tau}=\iota_{a0}(\delta_a,\delta_b,\dot u_1)-\beta_\text{rf} \sin \delta_a -\gamma\dot{\delta}_a\\
&\frac{d\dot{\delta}_b}{d\tau}=\iota_{b0}(\delta_a,\delta_b,\dot u_1)-\beta_\text{rf} \sin \delta_b -\gamma\dot{\delta}_b\\
&\frac{du_1}{d\tau}=\dot{u}_1\\
&\frac{d\dot{u}_1}{d\tau}=\ddot{u}_1(\dot \delta_a, \dot \delta_b,u_1 )
\end{align}
    
 \end{subequations}
Let's now examine the linear limit approximation for this problem. Consider the solutions in the following form  $\vec\delta=\vec\delta_\text{rf}(t)+\vec\delta_\text{dc}$, $\vec\delta_\text{rf}(t)=\hat{\vec\delta}_\text{rf}\exp(i\Omega \tau)$, and $u_1(t)=\hat u_1\exp(i\Omega \tau)$, where $\hat{\vec\delta}_\text{rf}=(\hat\delta_a,\hat\delta_b)$.  Substituting these expressions in the equation of motion Eq. (\ref{eqn:eom_con_corner2}) and Faraday's law, Eq. (\ref{eqn:faraday_center_corner2}), and rearranging the excitation to the left hand side of the equations, one can obtain a linear system in $(\hat{\delta}_a,\hat{\delta}_b, \hat u_1)$ :
\begin{align}
&\vec{\phi}_\text{dc}=\vec\delta_\text{dc}+\beta_\text{rf}\overleftrightarrow{\kappa}\sin\vec\delta_\text{dc}\nonumber\\    
&\begin{pmatrix}
    \phi_\text{rf}\\ \phi_\text{rf}\\ \dot\phi_\text{rf,cen}
\end{pmatrix}=
\overleftrightarrow\chi\cdot
\begin{pmatrix}
    \hat\delta_a\\ \hat\delta_b\\ \hat u_1
\end{pmatrix}
\label{eqn:alt_lin_2_corner_cpl}
\end{align}
where 
\begin{eqnarray*}
    \overleftrightarrow\chi=\hskip 20em \\
  \begin{pmatrix}
1 +\hat{\iota}_{\text{rf},a0} & \kappa_1\hat{\iota}_{\text{rf},a0}&i\Omega\frac{M_{\text{cen},a0}+L_{\text{cen},a1}}{L_\text{geo}}\hat{\iota}_{\text{rf},a0}
\\
\kappa_1\hat{\iota}_{\text{rf},b0} & 1 +\hat{\iota}_{\text{rf},a0} & i\Omega(\frac{L_{\text{cen},b0}+M_{\text{cen},b1}}{L_\text{geo}}\hat{\iota}_{\text{rf},b0}+1) \\
 -2i\lambda_\text{cov}\kappa_{\delta a}\Omega    &  2i\lambda_\text{cov}\kappa_{\delta b}\Omega&
    -2+2\lambda_\text{cov}\frac{L_{\text{cen},a1}-M_{\text{cen},b1}}{L_\text{geo}}\Omega^2
\end{pmatrix}^T
\end{eqnarray*}

and $\hat{\vec\iota}_\text{rf}=\beta_\text{rf}\cos\vec\delta_\text{dc}+i\gamma\Omega-\Omega^2$.
The resonance condition for the system is $\det(\overleftrightarrow\chi=0)$, which \textcolor{black}{is a sixth-order equation and} can be solved for $\Omega$ to obtain the eigen solutions to the two corner-coupled SQUIDs. Both the eigenfrequencies and the nonlinear numerical solutions have been obtained and \textcolor{black}{agree with the results} in Secs.  \ref{lin_lim_sol} and \ref{nonlin_mdl_2_corner_cpl} in the main text.\\

\section{A general model for a system of overlapping SQUIDs}
\label{app:general_volt_form_ovlp_SQUIDs}
As discussed in Sec. \ref{large_system}, the voltage formalism can be applied to any size of corner-coupled overlapping SQUIDs array. Here, we outline the procedure for setting up the equations of motion for a system of corner-coupled SQUIDs in any general geometry. The first step in studying the dynamics is to identify the independent variables, which are $\vec\delta$ from the junctions, $\vec V$ from the overlapping capacitors, and their time derivatives. We can then express the induced fluxes in the system as
\begin{eqnarray}
    \begin{pmatrix}
        \vec\Phi^\text{ind}_\text{SQUID}\\ \vec\Phi^\text{ind}_\text{non-SQUID}
    \end{pmatrix}=
    \overleftrightarrow L \overleftrightarrow I_\text{con}
    \begin{pmatrix}
        \vec I_\text{JJ} \\ c \dot{\vec V}
    \end{pmatrix}
\end{eqnarray}
where $\overleftrightarrow L$ is the inductance matrix whose $i,j$ the element describes the flux on the loop $i$ induced by the segment $j$,  and $\overleftrightarrow I_\text{con}$ is the matrix that expresses the currents in each branch in terms of junction currents $\vec I_\text{JJ}$ and the displacement currents through the capacitor nodes $c \dot{\vec V}$. Next, the same flux relations in the conventional RCSJ model (Sec. \ref{RCSJ_induct_theory}) are invoked here for each SQUID
\begin{eqnarray}
    \vec\Phi^\text{app}_\text{SQUID}-\vec\Phi^\text{ind}_\text{SQUID}=\frac{\Phi_0}{2\pi}\vec \delta
\end{eqnarray}
which can be solved for junction currents $\vec I_\text{JJ}(\vec\delta,\dot{\vec V})$. The non-SQUID loops, on the other hand, are described by Faraday's law in the following form

\begin{eqnarray}
    \dot{\vec\Phi}^\text{app}_\text{non-SQUID}-\dot{\vec\Phi}^\text{ind}_\text{non-SQUID}=\overleftrightarrow V_\text{con}
    \begin{pmatrix}
        \Phi_0/(2\pi) \dot{\vec\delta}\\\vec V
    \end{pmatrix}
    \label{eqn:faraday_general}
\end{eqnarray}
where $\overleftrightarrow V_\text{con}$ is the matrix that associates the junction voltages $\Phi_0/(2\pi) \dot{\vec\delta}$ and voltages across the capacitor nodes $\vec V$ to each Faraday loop. After substituting the expression  for $\vec I_\text{JJ}(\vec\delta,\dot{\vec V})$ into Eq. (\ref{eqn:faraday_general}), the second time derivatives of the capacitor voltages, $\ddot{\vec V}(\dot{\vec \delta},\vec V)$ are obtained. Consequently, we can formulate the final dimensionless  equations of motion in terms of $(\vec \delta,\dot{\vec \delta},\vec V, \dot{\vec V})$:
\begin{subequations}
\begin{eqnarray}
    \frac{d\vec\delta}{d\tau}=\dot{\vec\delta}\\
    \frac{d\dot{\vec\delta}}{d\tau}=\ddot{\vec\delta}=\vec\iota_\text{JJ}(\vec\delta,\dot{\vec u})-\beta_\text{rf}\sin(\vec\delta)-\gamma\dot{\vec\delta}\label{eqn:dddelta_general}\\
\frac{d\vec u}{d\tau}=\dot{\vec u}\\  
\frac{d\dot{\vec u}}{d\tau}=\ddot{\vec u}(\dot{\vec\delta},\vec u)\label{eqn:ddu_general}
\end{eqnarray}
\end{subequations}
\section{High power nonlinear numerical calculation }\label{app:high_pow_num_cal}
The nonlinear numerical solutions for the large systems of SQUIDs studied in Sec.\ref{large_system} at different applied rf flux amplitudes are summarized below. 
\begin{figure} 
    \centering
    \includegraphics[width=1\linewidth]{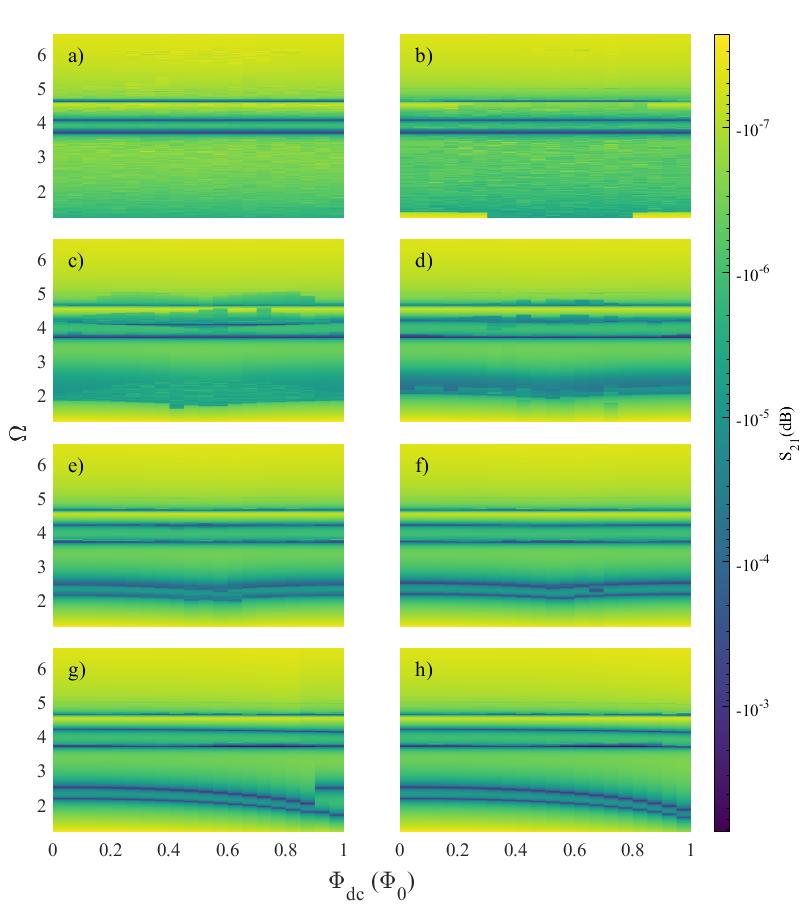}
    \caption{Calculated transmission of four corner-coupled SQUIDs as a function of dimensionless frequency $\Omega = \omega/\omega_\text{geo}$ and 
    as a function of dc flux swept from 0 to +1 $\Phi_0$. The solutions are obtained at different applied rf magnetic flux amplitudes: $[2.1,\ 0.66,\ 0.21,\ 0.066,\ 0.021,\ 0.0066,\ 0.0021,\ 0.00066]\ \Phi_0$ in panels a) through h).}
    \label{fig:4cnr_cpl_nlin_sol_diffrf}
\end{figure}
\begin{figure}
    \centering
    \includegraphics[width=1\linewidth]{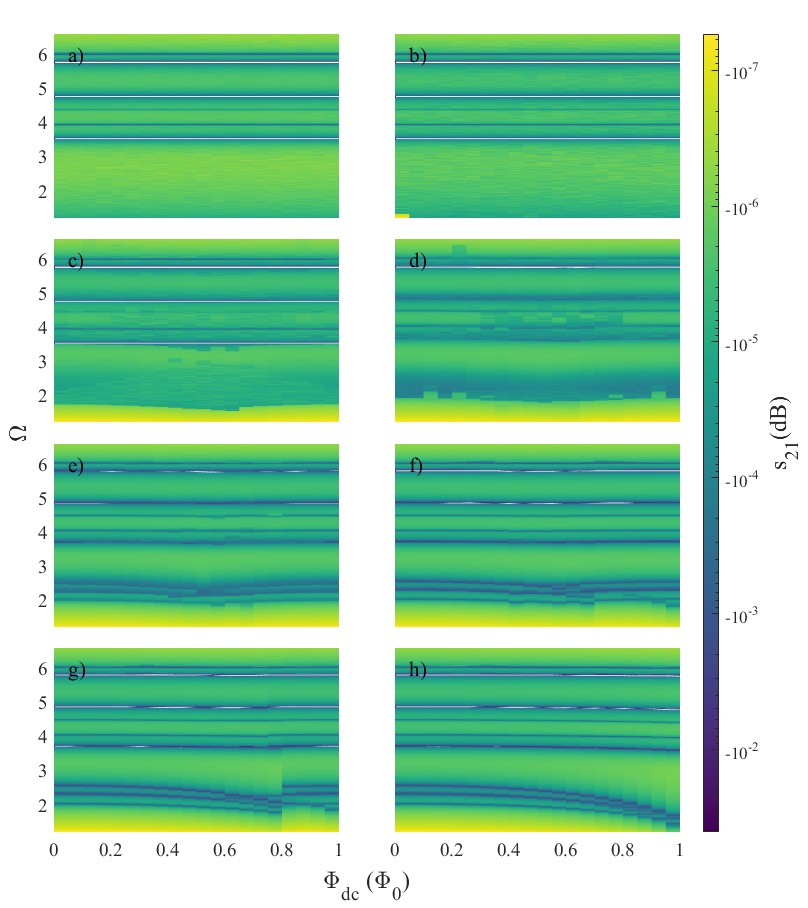}
    \caption{Calculated transmission of the $2\times2\times2$ system of  SQUIDs as a function of dimensionless frequency $\Omega = \omega/\omega_\text{geo}$ and as a function of dc flux swept from 0 to +1 $\Phi_0$. The solutions are obtained at different applied rf magnetic flux amplitudes: $[2.1,\ 0.66,\ 0.21,\ 0.066,\ 0.021,\ 0.0066,\ 0.0021,\ 0.00066]\ \Phi_0$ in panels a) through h).}
    \label{fig:2by2by2_nlin_sol_diffrf}
\end{figure}
There exists a clear distinction between the modes above $\Omega=3$ and the modes below in terms of their rf flux dependence in Fig.\ref{fig:4cnr_cpl_nlin_sol_diffrf} and \ref{fig:2by2by2_nlin_sol_diffrf}. The higher frequency modes are  insensitive to the applied flux, corresponding to the partial loop modes, while the lower frequency modes' dc flux tunabilities are suppressed at high power, typical for SQUID modes. This behavior is also observed in the experimental data on the $12\times 12\times 2$ SQUID metamaterial in Fig. \ref{fig:exp_2layer_array_rfdep}.
\
\\\\\\\\\\
\bibliography{WTD_new}

\end{document}